\documentclass[twocolumn,showpacs,preprintnumbers,amsmath,amssymb,aps,prl,superscriptaddress]{revtex4-1}
\makeatletter

\newcommand{\Rmnum}[1]{\expandafter\@slowromancap\romannumeral #1@}
\makeatother
\usepackage[colorlinks,linkcolor=blue,anchorcolor=blue,citecolor=blue,urlcolor=blue,breaklinks=true]{hyperref}

\usepackage{epsfig}
\usepackage{natbib}
\usepackage{epstopdf}
\usepackage{mathrsfs}
\usepackage{slashed}
\usepackage{amsmath}
\usepackage{verbatim}
\usepackage{graphicx}
\usepackage{amssymb}
\usepackage{psfrag}
\usepackage{array}
\usepackage{lipsum}
\usepackage{float}

\newcommand{\pT}{p_{\scriptscriptstyle T}}
\newcommand{\pTf}{p_{\scriptscriptstyle T}}
\newcommand{\pTi}{p_{\scriptscriptstyle T}^{\rm \tiny initial}}

\begin{document}

\title{Jet tomography in heavy ion collisions with deep learning}
\author{Yi-Lun Du}
\email{yilun.du@uib.no}
\affiliation{Department of Physics and Technology, University of Bergen, Postboks 7803, 5020 Bergen, Norway}
\author{Daniel Pablos}
\email{daniel.pablos.alfonso@to.infn.it}
\affiliation{INFN, Sezione di Torino, via Pietro Giuria 1, I-10125 Torino, Italy}
\author{Konrad Tywoniuk}
\email{konrad.tywoniuk@uib.no}
\affiliation{Department of Physics and Technology, University of Bergen, Postboks 7803, 5020 Bergen, Norway}
\begin{abstract}
Deep learning techniques have the power to identify the degree of modification of high energy jets traversing deconfined QCD matter on a jet-by-jet basis. Such knowledge allows us to study jets based on their initial, rather than final energy. We show how this new technique provides unique access to the genuine configuration profile of jets over the transverse plane of the nuclear collision, both with respect to their production point and their orientation. Effectively removing the selection biases induced by final-state interactions, one can in this way analyse the potential azimuthal anisotropies of jet production associated to initial-state effects.
Additionally, we demonstrate the capability of our new method to locate with precision the production point of a dijet pair in the nuclear overlap region, in what constitutes an important step forward towards the long term quest of using jets as tomographic probes of the quark-gluon plasma.
\end{abstract}

\maketitle


\paragraph{Introduction.} 
Jets are collimated sprays of hadrons that are produced in hard QCD processes in high-energy particle collisions \cite{Salam:2009jx,Larkoski:2017jix,Marzani:2019hun}. Within the context of heavy-ion collisions, they are witnesses to the creation of deconfined QCD matter, known as the quark-gluon plasma (QGP), which
behaves very close to a perfect liquid
\cite{ackermann2001elliptic,aamodt2010elliptic,aamodt2011higher}. During their passage through this medium, partonic jet modes are subject to momentum diffusion and energy loss by the radiation of soft quanta towards large angles, a phenomenon known as jet quenching \cite{dEnterria:2009xfs,Majumder:2010qh,Mehtar-Tani:2013pia,Blaizot:2015lma}. Key information about the medium is contained in the detailed modification of these hard probes, turning them into essential tools on which tremendous theoretical and experimental effort is being devoted \cite{Abelev:2013kqa,Adam:2015ewa,aad2015measurements,Aaboud:2017eww,aaboud2019measurement,Acharya:2019jyg}.

Using jets as differential probes of the spatio-temporal structure of the QGP created in heavy-ion collisions, aka \emph{jet tomography}, is a long-standing goal \cite{Wang:2002ri,Renk:2006qg,Zhang:2007ja,Zhang:2009rn,He:2020iow,PhysRevLett.127.082301}. On a jet-by-jet basis it is evident that the modifications induced by the medium follow from the local properties sampled along the jet trajectory from the hard production point out to the detector. The ability to unambiguously gauge the effect from the QGP on this level would lead to unprecedented precision in determining local properties of the fluid, including flow \cite{Armesto:2004vz,Sadofyev:2021ohn}, path-length dependence of modifications \cite{Betz:2014cza} and the possibility of observing deconfined quasi-particle degrees of freedom in the QGP \cite{DEramo:2018eoy,Barata:2020rdn,Harris:2020ijy}. Nonetheless, tomographic analyses on the level of inclusive jet populations have been hindered by intrinsic biases that accentuate samples experiencing small modifications over samples that are strongly affected~\cite{Baier:2001yt}. Such biases arise due to the steeply falling spectrum of the jet initiator transverse momenta and strongly distort the magnitude of medium effects, e.g. the in-medium path-length distribution of surviving jets. 


In this Letter we propose a technique, based on deep learning, that mitigates these bias effects and results in a better control of the path-length traversed by individual jets based on their level of modification. Given a measured jet at $\pTf$ and cone size $R$, the procedure allows us to estimate with reasonable accuracy the transverse momentum $p_T^{\rm initial}$ 
the jet would have had, had it not interacted with a medium, see \cite{Du:2020pmp} for further details on how to establish such a correspondence.
The technique uses only the information of the hadrons that are contained in the reconstructed jet and is easily adaptable to other model studies.

Having at hand an estimate of how much energy an individual jet has lost is a powerful tool that allows for many interesting applications \cite{Du:2020pmp}. 
Here, we demonstrate the usefulness of our approach to tomographic applications in two concrete examples. The first deals with reconstructing the true distribution of path-lengths that  jets experience, eliminating the effects of ``surface bias'' \cite{Dainese:2004te,Zhang:2007ja,Zhang:2009rn} and revealing the potential contributions to jet azimuthal anisotropy that do not stem from final-state interactions.
The second application combines the extraction of the lost energy with accessible knowledge about the orientation of the jet with respect to the event plane of the collision, as determined by the dominant azimuthal harmonic $v_2$ of the particle distribution. This allows to constrain the path length dependence separately for jets traveling parallel and transverse to the event plane of the collisions, refining the path to experimentally pinning down the original production point of a dijet pair. We expect this new development to importantly contribute to the set of tools aimed at the exploitation of energetic jets as tomographic probes of the QGP.

\paragraph{Quantifying energy loss jet-by-jet.}
In vacuum, high energy partons produced in a hard QCD collision relax their large virtuality down to the hadronization scale via successive splittings. 
The description of these processes is well controlled both from theory \cite{Dasgupta:2014yra,Dasgupta:2016bnd,Kang:2016mcy} and within Monte Carlo parton shower generators \cite{Buckley:2019kjt} through the appropriate implementation of the DGLAP evolution equations.
In the medium, the presence of a significant phase space for vacuum-like emissions, occurring before any medium induced modifications have had the time to develop, has been firmly established
\cite{Mehtar-Tani:2017web,Caucal:2018dla,Dominguez:2019ges}. The presence of this vacuum-like phase space, which strongly impacts the amount of jet energy loss based on the vacuum-set scales governing jet activity (multiplicity, i.e. number of energy loss sources), is in fact what allows us to understand a great number of jet quenching observables, such as the relative suppression between jets and hadrons \cite{Mehtar-Tani:2017web,Casalderrey-Solana:2018wrw}, or the narrowing of the angular opening between the leading groomed subjets \cite{Casalderrey-Solana:2019ubu,Caucal:2019uvr}.
In the leading-logarithmic approximation, it is legitimate to assume that the in-cone emissions that belong to the vacuum-like dominated region
already define the energy that the jet would have had in the absence of the medium, which we call $\pTi$.
The presence of the medium alters this fairly developed structure, typically leading to energy loss due to transport of particles out of the jet cone and defining the jet energy in the medium, or simply $\pTf$.

Within this factorised picture, we can define what we call the energy loss ratio $\chi \equiv \pTf/\pTi$. 
In Ref.~\cite{Du:2020pmp} we describe the  matching procedure carried out at hadron level necessary to establish this connection between a quenched jet and its vacuum-like counterpart. In our previous and current work, we use the hybrid strong/weak coupling model \cite{casalderrey2015erratum,Casalderrey-Solana:2015vaa,Casalderrey-Solana:2016jvj} for the generation of the quenched jets.
We extract $\chi$ on a jet-by-jet basis by using jet images as inputs to a convolutional neural network (CNN), achieving a good degree of accuracy across a wide range in $\chi$ \footnote{\textcolor{black}{See the supplemental material for a preliminary check on the model dependence of the extraction of $\chi$.}}. We refer the interested readers to our previous paper \cite{Du:2020pmp} for further details on data pre-processing and software architecture (see also \cite{Apolinario:2021olp} for a complementary approach). 

\paragraph{Factoring out final-state effects.}
In nucleus-nucleus collisions, the production of hard processes can be described by the Glauber model \cite{biallas1976multiplicity,miller2007glauber}, where the rate of collisions is governed by the inelastic cross section of nucleon-nucleon scatterings and the density of nucleons are described by the Woods-Saxon distribution. 
The distribution of production points is naturally strongly correlated with the distribution of path-lengths experienced by the entire jet population. However, jets that experience final-state interactions will tend to be more modified, and experience, on average, more energy loss if they originate from production points deep within the nuclear overlap region rather than from the surface.
Therefore, selecting a jet population based on their \textit{final}, measured transverse momenta will bias the jet selection toward short path-lengths, and small energy losses, leading to a ``surface bias'' \cite{Dainese:2004te,Zhang:2007ja,Zhang:2009rn}. In contrast, focusing on the original jet population, or selecting jets according to their \textit{initial} transverse momentum, accessed with $\pTi =  \pTf/\chi$, should recover the true path-length distribution associated to the underlying nuclear overlap density.

\begin{figure}[tb!]
\centering
\includegraphics[width=0.48\textwidth]{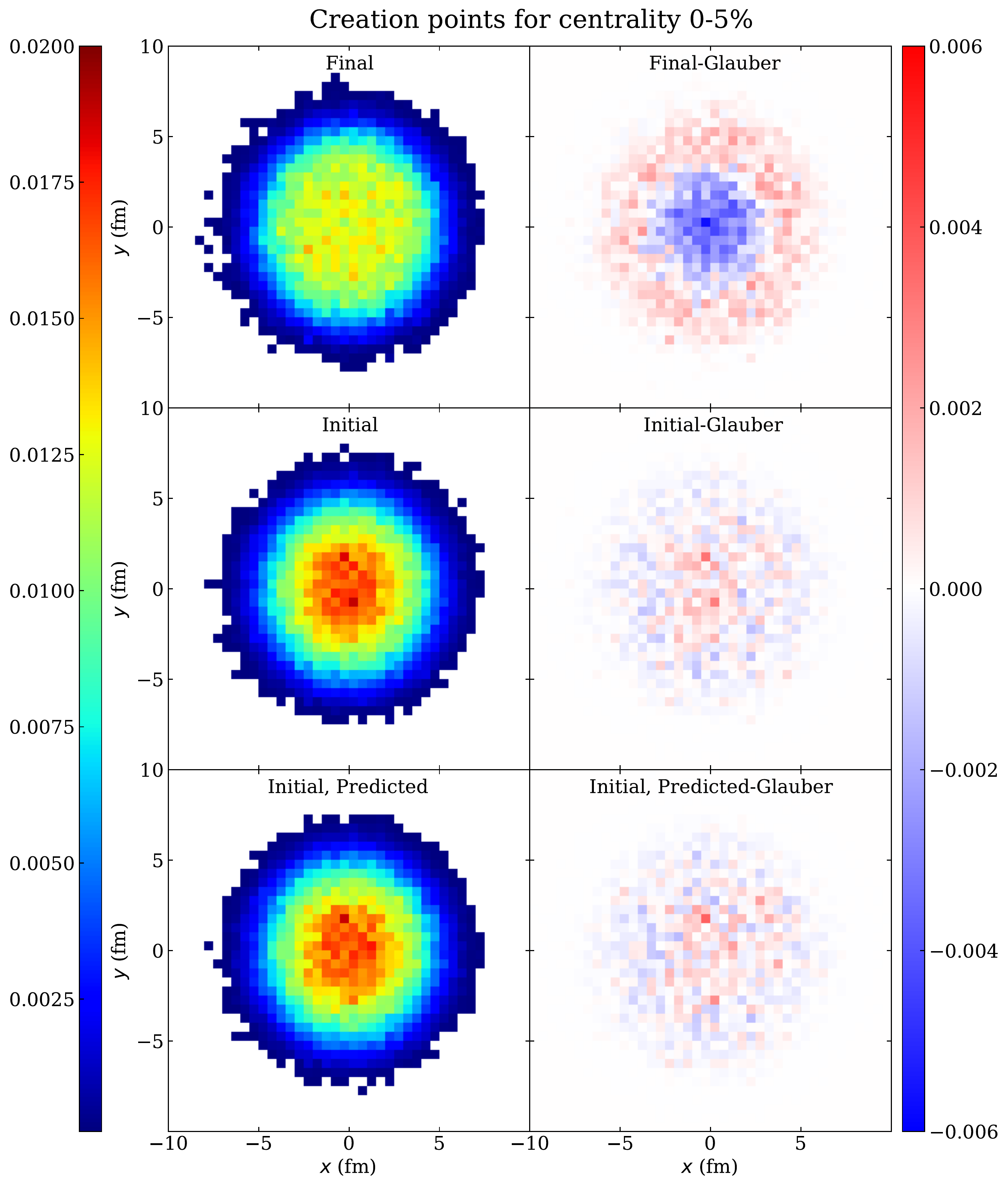} 
\caption{Left: Probability distribution of the production point in the transverse plane of a hard QCD process when using the FES setup, in the top, versus when IES setup is used, with true $\chi$ in the middle and with predicted $\chi^p$ in the bottom. Right: Difference of the results of the left column with respect to the distribution obtained by directly using the Glauber procedure.} 
\label{fig: creation points}
\end{figure}
In order to visualize these aspects, we generate dijet events at $\sqrt{s_{NN}}= 5.02$ TeV for PbPb collisions in the \mbox{0-5\%} centrality bin corresponding to around 700,000 samples of inclusive jets reconstructed with anti-$k_t$ \cite{Cacciari:2008gp} and radius parameter $R=0.4$ using FastJet \cite{Cacciari:2011ma}. In the left column of Fig.~\ref{fig: creation points} we show the production point density of the hard QCD processes in the transverse plane using three different jet selections. In the top left panel, we select jets with a measured momentum $\pT>200$ GeV and plot the location where they were produced (taken directly from the model). This selection, referred to as the \emph{final energy selection} (FES), is the only possible setup in experiments without the knowledge of $\chi$. Taking the difference with respect to the actual production point density using the Glauber model in the top right plot of Fig.~\ref{fig: creation points} we observe that, compared to the true geometrical distribution according to which the jets were produced, there is a relative absence of jets produced at the centre of the overlap region. 

With a good estimate of the energy loss ratio $\chi$ we can, however, perform a different jet selection. In the middle and bottom left plots of Fig.~\ref{fig: creation points} we show the jet production points (again, supplied by the model) for the so-called \emph{initial energy selection} (IES) with true, i.e. extracted directly from model data, and predicted, i.e. extracted by the trained CNN, $\chi$ respectively, where we only include those jets with $\pTi > 200$ GeV \footnote{Additionally, we demand that $\pTf>100$ GeV because this corresponds to the kinematical range where we have trained our neural network. We have also checked that our results do not depend on the precise cut on $\pTf$ as long as it is sufficiently below the cut of $\pTi$.}. 
Remarkably, the differences with respect to the Glauber distribution, shown in the middle and bottom right panels, display no sizeable deviation beyond random noise. \textcolor{black}{A detailed error analysis is presented in the supplementary material.} This demonstrates that, by employing IES, we are able to mitigate \textcolor{black}{almost all} final-state effects, such as medium-induced energy loss, and we obtain a jet population that reflects the true path-length distribution experienced in a heavy-ion collision.

\begin{figure}[t!]
\centering
\includegraphics[width=0.48\textwidth]{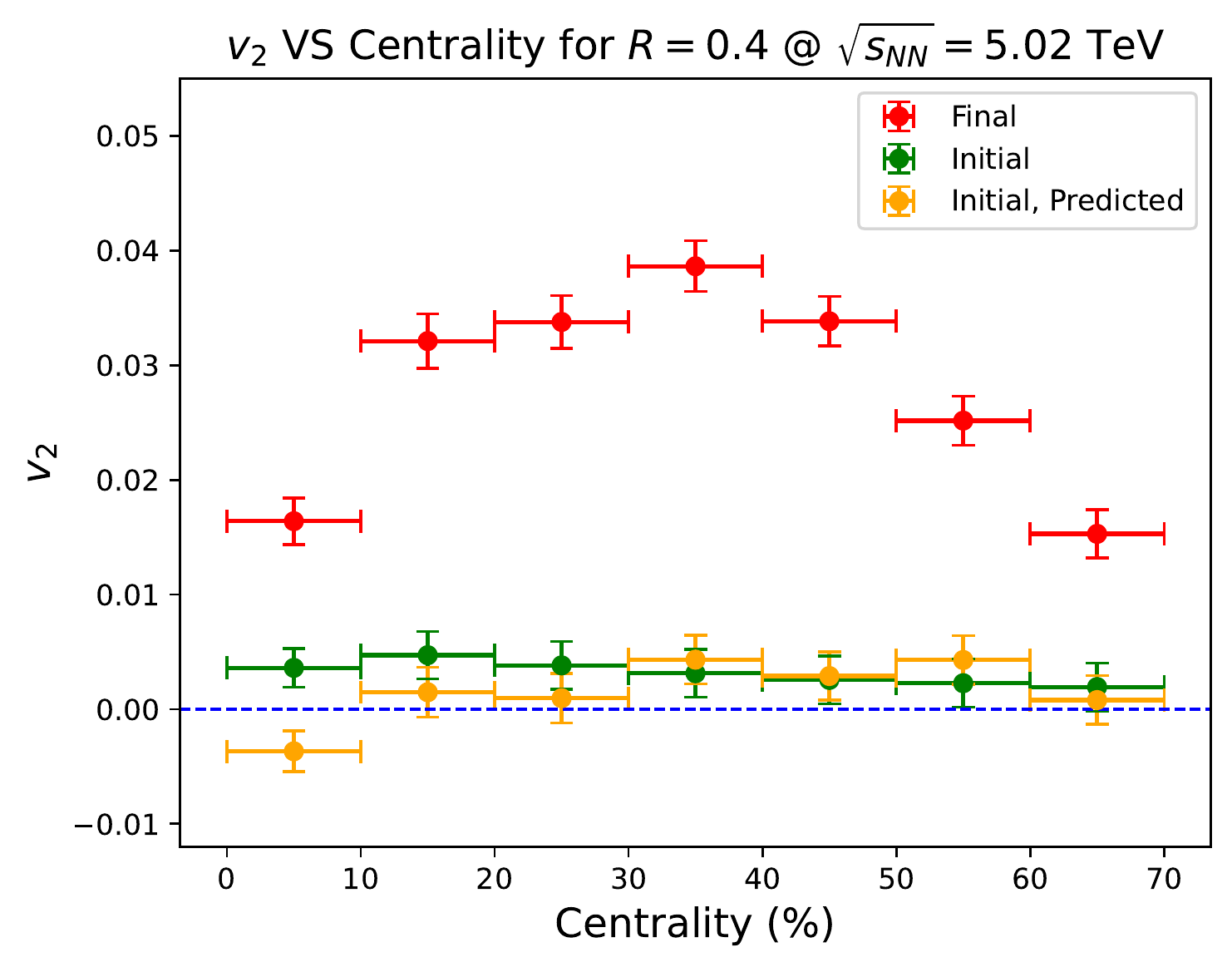}
\caption{Centrality dependence of $v_2$ for FES setup (red) and IES setup with true $\chi$ (green) and predicted $\chi^p$ (orange).
} 
\label{fig: V2_VS_centrality}
\end{figure}

Selection bias effects affect not only the creation point distribution in the transverse plane but also the jet orientation with respect to the event plane of the collision \cite{Wiedemann:2008zz}. In contrast to the production points shown in Fig.~\ref{fig: creation points}, the azimuthal distribution of particle production can be measured in experiments and quantified by the harmonic coefficients $v_n = \langle \exp[i\, n\phi] \rangle$, where $\phi$ is the azimuthal angle with respect to the event plane and the average is taken over all measured events. In particular, for high-$\pT$ probes, the second harmonic coefficient $v_2$ is given directly by $v_2 = \left\langle \big( p_{{\scriptscriptstyle T},x}^2-p_{{\scriptscriptstyle T},y}^2\big)\big/ \big( p_{{\scriptscriptstyle T},x}^2+p_{{\scriptscriptstyle T},y}^2\big) \right\rangle$.
These momentum anisotropies can emerge both due to initial-state correlations and final-state interactions~\cite{Petersen:2012qc,Greif:2017bnr,Nie:2019swk}. The former arise from quantum interference in the incoming nuclear wave-functions and dominate the observable signal of $v_2$ only at small multiplicities \cite{Giacalone:2020byk}. The latter are generally driven by the geometry of the collisions (the nuclear positions within the nuclei at a given impact parameter)~\cite{gyulassy2001high,wang2001jet}. 

We show results for jet $v_2$ from the hybrid strong/weak coupling model in Fig.~\ref{fig: V2_VS_centrality} in PbPb collisions at $\sqrt{s_{NN}}= 5.02$ TeV, using anti-$k_t$ and $R=0.4$, as a function of centrality. The red dots correspond to the obtained $v_2$ using FES for jets with measured $p_T>200$ GeV (we have checked that the results from the hybrid model reproduce experimental data on high-$p_T$ $v_2$ at $\sqrt{s_{NN}} = 2.76$ TeV \cite{aad2013measurement} very well; see the supplemental material for more details). 
As the nuclear overlap region becomes more and more anisotropic with increasing impact parameter (increasing the centrality class), final-state energy loss effects increasingly enhance the relative contribution of the less quenched jets that propagate along the short axis of the collision. Thus, $v_2^{\rm \tiny FES}$ is positive and grows with centrality. However, as the medium becomes smaller and colder at the most peripheral collisions, energy loss is reduced and $v_2^{\rm \tiny FES}$ consistently decreases~\cite{wang2001jet,gyulassy2001high}.

On the other hand, the green and orange points in Fig.~\ref{fig: V2_VS_centrality} correspond to the results for the IES procedure, using the true and predicted values of $\chi$, respectively, for jets with $\pTi > 200$ GeV.
By removing the selection bias effect we reveal the initial-state orientation of hard jets in our model which, by construction, is random and therefore $v^{\rm \tiny IES}_2$ is consistent with zero.
Remarkably, the agreement between the green and orange dots demonstrates that our algorithm, having been trained on jets in the 0-5\% centrality class, is generalizable across a wide range of centrality classes. 
We also note that our method would have yielded $v_2^{\rm IES} \neq 0$ and revealed any remaining anisotropy associated to other hypothesized mechanisms, such as quantum correlations in the initial wave functions~\cite{lappi2016tracing,mace2018hierarchy,PhysRevLett.122.172302, Albacete:2016gxu,gelis2019primordial} or other quantum interference effects~\cite{Blok:2017pui,Blok:2018xes}\footnote{In principle, given the absence of energy loss in colorless particles, the value of $v_2$ for high-$p_T$ photons \cite{Hamed:2014hta} or $Z^0$ bosons should also be sensitive to such initial-state induced anisotropy.}, \textcolor{black}{although currently known mechanisms are conjectured to average out in large systems due to combinatorial effects}. While such effects at high-$\pT$ are expected to be small within current models, finding evidence of such \textit{additional} anisotropies in nucleus-nucleus collisions would support the idea of a common, underlying contribution to collective behavior across different system sizes. 

\paragraph{Jet tomography of the QGP.}
\begin{figure*}[t!]
\includegraphics[width=0.95\textwidth]{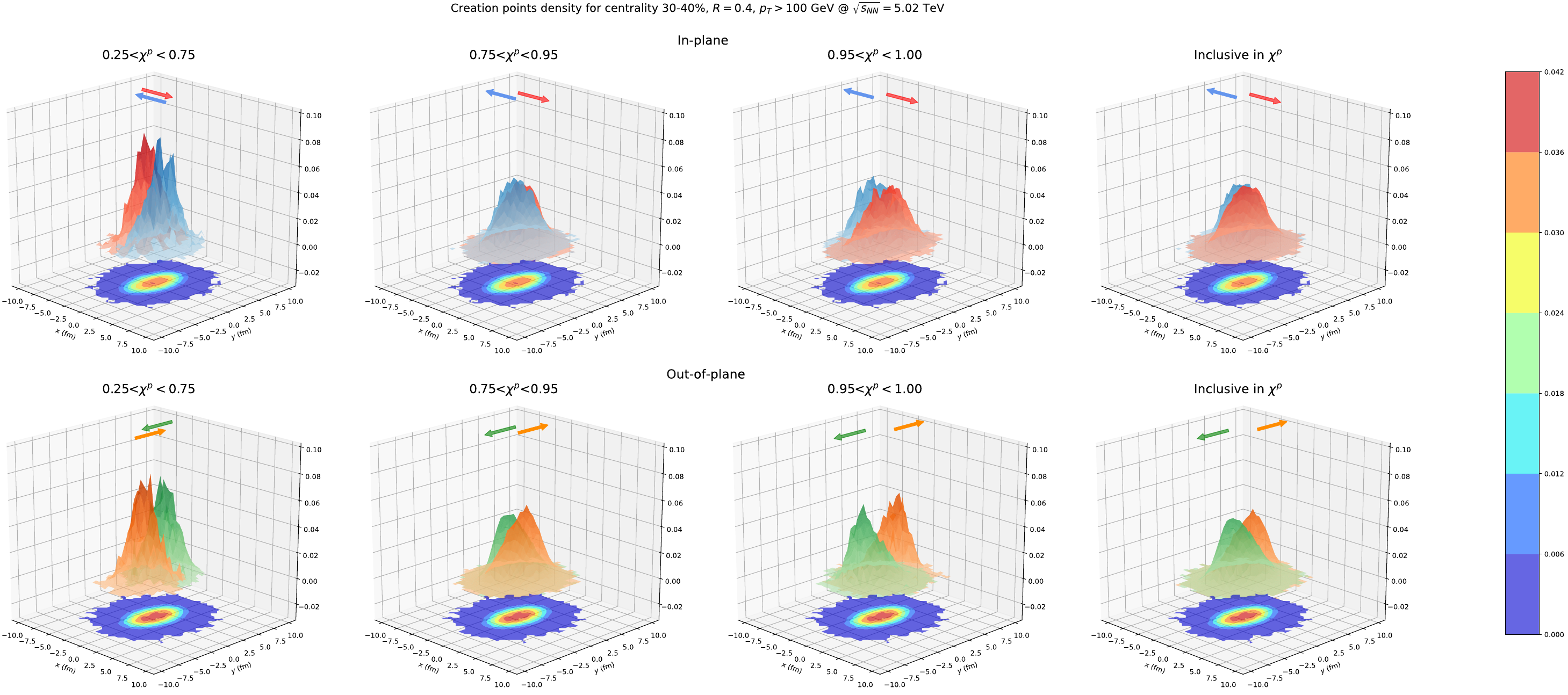}
\caption{Creation point distributions in the transverse plane for the jets in 30-40\% centrality and inclusive and sliced in different ranges of the predicted $\chi^p$ in four columns, respectively. The in-plane jets going left (blue) and right (red) are shown in the upper row and the out-of-plane jets going up (orange) and down (green) are shown in the lower row. The 2-D histogram in the bottom of each plot is the distribution of the inclusive in-plane (upper row) and out-of-plane (lower row) jets in this centrality.}
\label{fig: Tomography_centrality_30_40_in_out_plane_FES_chip} 
\end{figure*}
We now turn to the final application of our tomographic studies using deep learning. Having established that we can restore the true path-length distribution of jets above, we can further narrow down the path-length selection by choosing jets within a class of energy loss ratios, i.e. by choosing jets in a specific range of $\chi$, \textcolor{black}{similar in spirit to the uses of boson-jet selections \cite{Zhang:2009rn,PhysRevLett.127.082301} in which the boson energy serves as a proxy for the initial hard parton energy}. However, due to the many sources of fluctuations (from jet substructure fluctuations to fluctuations residing in the medium interactions), the correlation between $\chi$ and the path length is not as strong as one could have expected \cite{Milhano:2015mng,Du:2020pmp}. Notwithstanding, from simple geometrical considerations the path-length distribution, and consequently the possible production points in the nuclear overlap area, can be constricted further by additionally constricting the direction of the jet propagation with respect to the event plane of the collisions. Concretely, we will consider jets propagating in-plane, i.e. parallel to the event plane, and out-of-plane, i.e. transverse to the event plane. 

By combining our knowledge of the orientation of a jet with respect to the event plane with the degree of energy loss $\chi$, we present the localization of the production point of a given jet with a new level of precision. In Fig.~\ref{fig: Tomography_centrality_30_40_in_out_plane_FES_chip} we show results for around 900,000 jets, generated at $\sqrt{s_{NN}}= 5.02$ TeV for PbPb collisions at 30-40\% centrality and reconstructed using anti-$k_t$ and $R=0.4$. In the upper (lower) row we have selected jets that are propagating in-plane (out-of-plane), which means they are approximately oriented along the short (long) axis of the nuclear overlap region. This corresponds to jets with distinctly positive (negative) $v_2$. In the bottom of each sub-figure in the upper (lower) row we also plot the average nuclear overlap density, represented by the in-plane (out-of-plane) jet production point distribution for FES inclusive in $\chi$,
which has been rotated so that the event plane, and impact parameter vector, points along the $x$-direction. 
We can further slice the selection depending on which sense the jet is propagating: either left (in blue) or right (in red) for the in-plane jets, and either up (in orange) or down (in green) for the out-of-plane jets. The histograms in each of the first three columns display the creation point density for jets belonging to a given range of the predicted energy loss ratio $\chi^p$. Finally, the fourth column shows the results inclusive in predicted $\chi^p$ and corresponds to the production point distributions if we had no knowledge of the degree of in-medium modification. 

Even though there is some degree of separation of the production point distributions for the $\chi$-inclusive jet selection, a large degree of overlap can be noticed. This situation changes radically by using our knowledge of predicted $\chi^p$ on a jet-by-jet basis \footnote{A reasonable estimate of the amount of medium induced modification can also be obtained from the ratio between the jet $p_T$ and that of a recoiling colorless trigger boson, although the correlation is not tight even in vacuum due to sizeable out-of-cone radiation \cite{Zhang:2009rn}.}. The third column in Fig.~\ref{fig: Tomography_centrality_30_40_in_out_plane_FES_chip} shows results for fairly unquenched jets, with $0.95<\chi^p<1$. In order to belong to this class, a jet propagating upwards (focusing first on the out-of-plane jets in the lower row) has to have traversed merely a short distance through the QGP and therefore its production point is predominantly localized in the \textit{upper} part of the overlap region (and vice versa for a jet propagating downwards). This reasoning also applies, in reverse, for jets belonging to the very quenched class, with $0.25<\chi^p<0.75$, displayed in the leftmost column. In this case, a very quenched jet propagating upwards will have needed to traverse a long distance in the QGP, or analogously through a hot region, and consequently its production point will instead be predominantly localized in the \textit{lower} hemisphere (and vice versa for the downward propagating case). Obviously, analogous arguments also apply for the in-plane cases plotted in the upper row of Fig.~\ref{fig: Tomography_centrality_30_40_in_out_plane_FES_chip}. The second column shows the notably overlapping transition region that bridges the gap between the fairly unquenched and very quenched classes of columns one and three, respectively.

\paragraph{Conclusions.} 
In this Letter, we demonstrate the power of deep learning techniques to pin down the genuine nuclear density distributions that affect high-$\pT$ jet production in the transverse plane and the initial-state jet azimuthal anisotropies, quantifiable through the elliptic flow coefficient $v_2$, by learning the amount of energy loss $\chi$ on a jet-by-jet basis. In both cases, the final-state effects induced by the selection bias can be removed to a large extent by selecting a jet population according to their initial transverse momenta.
\textcolor{black}{It would be very interesting to assess the performance and $p_T$ reach of our technique when compared to other methods of obtaining relatively unbiased distributions, such as the quantile procedure \cite{Brewer:2018dfs,Takacs:2021bpv} or the use of boson-jet systems \cite{Zhang:2009rn,Casalderrey-Solana:2015vaa,Takacs:2021bpv}.}
We have argued that extracting the possible initial-state anisotropies in nucleus-nucleus collisions can serve to clarify the origin of the high-$\pT$ $v_2$ measured in small systems, which currently seems in strong conflict with an explanation based solely on energy loss physics. Furthermore, by selecting jets according to the sign of $v_2$, we have shown the capability of our method to locate \textcolor{black}{with precision} the jet creation point in the transverse plane, representing a significant development towards using jets as tomographic probes of the QGP. 
The interplay between the jet and the local properties of the medium, such as the local hydrodynamic flow \cite{Armesto:2004vz,Yan:2017rku,Tachibana:2020mtb,Casalderrey-Solana:2020rsj,Sadofyev:2021ohn} or temperature and density gradients~\cite{He:2020iow,PhysRevLett.127.082301,Sadofyev:2021ohn}, which determine preferred directions and deformed radiation spectra for the soft quanta emitted from the jet, 
could also be used to greatly improve the prediction performance of $\chi$ and even allow for a direct extraction of the traversed length in the QGP. This is a challenging task that will be tackled in future work.

\emph{Acknowledgements}. This work is supported by the Trond Mohn Foundation under Grant No. BFS2018REK01 and the University of Bergen. Y.D. thanks the support from the Norwegian e-infrastructure UNINETT Sigma2 for the data storage and HPC resources with Project Nos. NS9753K and NN9753K. D.P. has received funding from the European Union's Horizon 2020 research and innovation program under the Marie Skłodowska-Curie grant agreement No. 754496.

\bibliographystyle{apsrev4-1}
\bibliography{duyl}

\begin{thebibliography}{74}%
\makeatletter
\providecommand \@ifxundefined [1]{%
 \@ifx{#1\undefined}
}%
\providecommand \@ifnum [1]{%
 \ifnum #1\expandafter \@firstoftwo
 \else \expandafter \@secondoftwo
 \fi
}%
\providecommand \@ifx [1]{%
 \ifx #1\expandafter \@firstoftwo
 \else \expandafter \@secondoftwo
 \fi
}%
\providecommand \natexlab [1]{#1}%
\providecommand \enquote  [1]{``#1''}%
\providecommand \bibnamefont  [1]{#1}%
\providecommand \bibfnamefont [1]{#1}%
\providecommand \citenamefont [1]{#1}%
\providecommand \href@noop [0]{\@secondoftwo}%
\providecommand \href [0]{\begingroup \@sanitize@url \@href}%
\providecommand \@href[1]{\@@startlink{#1}\@@href}%
\providecommand \@@href[1]{\endgroup#1\@@endlink}%
\providecommand \@sanitize@url [0]{\catcode `\\12\catcode `\$12\catcode
  `\&12\catcode `\#12\catcode `\^12\catcode `\_12\catcode `\%12\relax}%
\providecommand \@@startlink[1]{}%
\providecommand \@@endlink[0]{}%
\providecommand \url  [0]{\begingroup\@sanitize@url \@url }%
\providecommand \@url [1]{\endgroup\@href {#1}{\urlprefix }}%
\providecommand \urlprefix  [0]{URL }%
\providecommand \Eprint [0]{\href }%
\providecommand \doibase [0]{http://dx.doi.org/}%
\providecommand \selectlanguage [0]{\@gobble}%
\providecommand \bibinfo  [0]{\@secondoftwo}%
\providecommand \bibfield  [0]{\@secondoftwo}%
\providecommand \translation [1]{[#1]}%
\providecommand \BibitemOpen [0]{}%
\providecommand \bibitemStop [0]{}%
\providecommand \bibitemNoStop [0]{.\EOS\space}%
\providecommand \EOS [0]{\spacefactor3000\relax}%
\providecommand \BibitemShut  [1]{\csname bibitem#1\endcsname}%
\let\auto@bib@innerbib\@empty
\bibitem [{\citenamefont {Salam}(2010)}]{Salam:2009jx}%
  \BibitemOpen
  \bibfield  {author} {\bibinfo {author} {\bibfnamefont {G.~P.}\ \bibnamefont
  {Salam}},\ }\href {\doibase 10.1140/epjc/s10052-010-1314-6} {\bibfield
  {journal} {\bibinfo  {journal} {Eur. Phys. J. C}\ }\textbf {\bibinfo {volume}
  {67}},\ \bibinfo {pages} {637} (\bibinfo {year} {2010})},\ \Eprint
  {http://arxiv.org/abs/0906.1833} {arXiv:0906.1833 [hep-ph]} \BibitemShut
  {NoStop}%
\bibitem [{\citenamefont {Larkoski}\ \emph {et~al.}(2020)\citenamefont
  {Larkoski}, \citenamefont {Moult},\ and\ \citenamefont
  {Nachman}}]{Larkoski:2017jix}%
  \BibitemOpen
  \bibfield  {author} {\bibinfo {author} {\bibfnamefont {A.~J.}\ \bibnamefont
  {Larkoski}}, \bibinfo {author} {\bibfnamefont {I.}~\bibnamefont {Moult}}, \
  and\ \bibinfo {author} {\bibfnamefont {B.}~\bibnamefont {Nachman}},\ }\href
  {\doibase 10.1016/j.physrep.2019.11.001} {\bibfield  {journal} {\bibinfo
  {journal} {Phys. Rept.}\ }\textbf {\bibinfo {volume} {841}},\ \bibinfo
  {pages} {1} (\bibinfo {year} {2020})},\ \Eprint
  {http://arxiv.org/abs/1709.04464} {arXiv:1709.04464 [hep-ph]} \BibitemShut
  {NoStop}%
\bibitem [{\citenamefont {Marzani}\ \emph {et~al.}(2019)\citenamefont
  {Marzani}, \citenamefont {Soyez},\ and\ \citenamefont
  {Spannowsky}}]{Marzani:2019hun}%
  \BibitemOpen
  \bibfield  {author} {\bibinfo {author} {\bibfnamefont {S.}~\bibnamefont
  {Marzani}}, \bibinfo {author} {\bibfnamefont {G.}~\bibnamefont {Soyez}}, \
  and\ \bibinfo {author} {\bibfnamefont {M.}~\bibnamefont {Spannowsky}},\
  }\href {\doibase 10.1007/978-3-030-15709-8} {\emph {\bibinfo {title}
  {{Looking inside jets: an introduction to jet substructure and boosted-object
  phenomenology}}}},\ Vol.\ \bibinfo {volume} {958}\ (\bibinfo  {publisher}
  {Springer},\ \bibinfo {year} {2019})\ \Eprint
  {http://arxiv.org/abs/1901.10342} {arXiv:1901.10342 [hep-ph]} \BibitemShut
  {NoStop}%
\bibitem [{\citenamefont {Ackermann}\ \emph {et~al.}(2001)\citenamefont
  {Ackermann}, \citenamefont {Adams}, \citenamefont {Adler}, \citenamefont
  {Ahammed}, \citenamefont {Ahmad}, \citenamefont {Allgower}, \citenamefont
  {Amsbaugh}, \citenamefont {Anderson}, \citenamefont {Anderssen},
  \citenamefont {Arnesen} \emph {et~al.}}]{ackermann2001elliptic}%
  \BibitemOpen
  \bibfield  {author} {\bibinfo {author} {\bibfnamefont {K.}~\bibnamefont
  {Ackermann}}, \bibinfo {author} {\bibfnamefont {N.}~\bibnamefont {Adams}},
  \bibinfo {author} {\bibfnamefont {C.}~\bibnamefont {Adler}}, \bibinfo
  {author} {\bibfnamefont {Z.}~\bibnamefont {Ahammed}}, \bibinfo {author}
  {\bibfnamefont {S.}~\bibnamefont {Ahmad}}, \bibinfo {author} {\bibfnamefont
  {C.}~\bibnamefont {Allgower}}, \bibinfo {author} {\bibfnamefont
  {J.}~\bibnamefont {Amsbaugh}}, \bibinfo {author} {\bibfnamefont
  {M.}~\bibnamefont {Anderson}}, \bibinfo {author} {\bibfnamefont
  {E.}~\bibnamefont {Anderssen}}, \bibinfo {author} {\bibfnamefont
  {H.}~\bibnamefont {Arnesen}},  \emph {et~al.},\ }\href@noop {} {\bibfield
  {journal} {\bibinfo  {journal} {Phys. Rev. Lett}\ }\textbf {\bibinfo {volume}
  {86}},\ \bibinfo {pages} {402} (\bibinfo {year} {2001})}\BibitemShut
  {NoStop}%
\bibitem [{\citenamefont {Aamodt}\ \emph {et~al.}(2010)\citenamefont {Aamodt},
  \citenamefont {Abelev}, \citenamefont {Quintana}, \citenamefont {Adamova},
  \citenamefont {Adare}, \citenamefont {Aggarwal}, \citenamefont {Rinella},
  \citenamefont {Agocs}, \citenamefont {Salazar}, \citenamefont {Ahammed} \emph
  {et~al.}}]{aamodt2010elliptic}%
  \BibitemOpen
  \bibfield  {author} {\bibinfo {author} {\bibfnamefont {K.}~\bibnamefont
  {Aamodt}}, \bibinfo {author} {\bibfnamefont {B.}~\bibnamefont {Abelev}},
  \bibinfo {author} {\bibfnamefont {A.~A.}\ \bibnamefont {Quintana}}, \bibinfo
  {author} {\bibfnamefont {D.}~\bibnamefont {Adamova}}, \bibinfo {author}
  {\bibfnamefont {A.}~\bibnamefont {Adare}}, \bibinfo {author} {\bibfnamefont
  {M.}~\bibnamefont {Aggarwal}}, \bibinfo {author} {\bibfnamefont {G.~A.}\
  \bibnamefont {Rinella}}, \bibinfo {author} {\bibfnamefont {A.}~\bibnamefont
  {Agocs}}, \bibinfo {author} {\bibfnamefont {S.~A.}\ \bibnamefont {Salazar}},
  \bibinfo {author} {\bibfnamefont {Z.}~\bibnamefont {Ahammed}},  \emph
  {et~al.},\ }\href@noop {} {\bibfield  {journal} {\bibinfo  {journal} {Phys.
  Rev. Lett}\ }\textbf {\bibinfo {volume} {105}},\ \bibinfo {pages} {252302}
  (\bibinfo {year} {2010})}\BibitemShut {NoStop}%
\bibitem [{\citenamefont {Aamodt}\ \emph {et~al.}(2011)\citenamefont {Aamodt},
  \citenamefont {Abelev}, \citenamefont {Quintana}, \citenamefont {Adamova},
  \citenamefont {Adare}, \citenamefont {Aggarwal}, \citenamefont {Rinella},
  \citenamefont {Agocs}, \citenamefont {Agostinelli}, \citenamefont {Salazar}
  \emph {et~al.}}]{aamodt2011higher}%
  \BibitemOpen
  \bibfield  {author} {\bibinfo {author} {\bibfnamefont {K.}~\bibnamefont
  {Aamodt}}, \bibinfo {author} {\bibfnamefont {B.}~\bibnamefont {Abelev}},
  \bibinfo {author} {\bibfnamefont {A.~A.}\ \bibnamefont {Quintana}}, \bibinfo
  {author} {\bibfnamefont {D.}~\bibnamefont {Adamova}}, \bibinfo {author}
  {\bibfnamefont {A.}~\bibnamefont {Adare}}, \bibinfo {author} {\bibfnamefont
  {M.}~\bibnamefont {Aggarwal}}, \bibinfo {author} {\bibfnamefont {G.~A.}\
  \bibnamefont {Rinella}}, \bibinfo {author} {\bibfnamefont {A.}~\bibnamefont
  {Agocs}}, \bibinfo {author} {\bibfnamefont {A.}~\bibnamefont {Agostinelli}},
  \bibinfo {author} {\bibfnamefont {S.~A.}\ \bibnamefont {Salazar}},  \emph
  {et~al.},\ }\href@noop {} {\bibfield  {journal} {\bibinfo  {journal} {Phys.
  Rev. Lett}\ }\textbf {\bibinfo {volume} {107}},\ \bibinfo {pages} {032301}
  (\bibinfo {year} {2011})}\BibitemShut {NoStop}%
\bibitem [{\citenamefont {d'Enterria}(2010)}]{dEnterria:2009xfs}%
  \BibitemOpen
  \bibfield  {author} {\bibinfo {author} {\bibfnamefont {D.}~\bibnamefont
  {d'Enterria}},\ }\href {\doibase 10.1007/978-3-642-01539-7_16} {\bibfield
  {journal} {\bibinfo  {journal} {Landolt-Bornstein}\ }\textbf {\bibinfo
  {volume} {23}},\ \bibinfo {pages} {471} (\bibinfo {year} {2010})},\ \Eprint
  {http://arxiv.org/abs/0902.2011} {arXiv:0902.2011 [nucl-ex]} \BibitemShut
  {NoStop}%
\bibitem [{\citenamefont {Majumder}\ and\ \citenamefont
  {Van~Leeuwen}(2011)}]{Majumder:2010qh}%
  \BibitemOpen
  \bibfield  {author} {\bibinfo {author} {\bibfnamefont {A.}~\bibnamefont
  {Majumder}}\ and\ \bibinfo {author} {\bibfnamefont {M.}~\bibnamefont
  {Van~Leeuwen}},\ }\href {\doibase 10.1016/j.ppnp.2010.09.001} {\bibfield
  {journal} {\bibinfo  {journal} {Prog. Part. Nucl. Phys.}\ }\textbf {\bibinfo
  {volume} {66}},\ \bibinfo {pages} {41} (\bibinfo {year} {2011})},\ \Eprint
  {http://arxiv.org/abs/1002.2206} {arXiv:1002.2206 [hep-ph]} \BibitemShut
  {NoStop}%
\bibitem [{\citenamefont {Mehtar-Tani}\ \emph {et~al.}(2013)\citenamefont
  {Mehtar-Tani}, \citenamefont {Milhano},\ and\ \citenamefont
  {Tywoniuk}}]{Mehtar-Tani:2013pia}%
  \BibitemOpen
  \bibfield  {author} {\bibinfo {author} {\bibfnamefont {Y.}~\bibnamefont
  {Mehtar-Tani}}, \bibinfo {author} {\bibfnamefont {J.~G.}\ \bibnamefont
  {Milhano}}, \ and\ \bibinfo {author} {\bibfnamefont {K.}~\bibnamefont
  {Tywoniuk}},\ }\href {\doibase 10.1142/S0217751X13400137} {\bibfield
  {journal} {\bibinfo  {journal} {Int. J. Mod. Phys. A}\ }\textbf {\bibinfo
  {volume} {28}},\ \bibinfo {pages} {1340013} (\bibinfo {year} {2013})},\
  \Eprint {http://arxiv.org/abs/1302.2579} {arXiv:1302.2579 [hep-ph]}
  \BibitemShut {NoStop}%
\bibitem [{\citenamefont {Blaizot}\ and\ \citenamefont
  {Mehtar-Tani}(2015)}]{Blaizot:2015lma}%
  \BibitemOpen
  \bibfield  {author} {\bibinfo {author} {\bibfnamefont {J.-P.}\ \bibnamefont
  {Blaizot}}\ and\ \bibinfo {author} {\bibfnamefont {Y.}~\bibnamefont
  {Mehtar-Tani}},\ }\href {\doibase 10.1142/S021830131530012X} {\bibfield
  {journal} {\bibinfo  {journal} {Int. J. Mod. Phys. E}\ }\textbf {\bibinfo
  {volume} {24}},\ \bibinfo {pages} {1530012} (\bibinfo {year} {2015})},\
  \Eprint {http://arxiv.org/abs/1503.05958} {arXiv:1503.05958 [hep-ph]}
  \BibitemShut {NoStop}%
\bibitem [{\citenamefont {Abelev}\ \emph {et~al.}(2014)\citenamefont {Abelev}
  \emph {et~al.}}]{Abelev:2013kqa}%
  \BibitemOpen
  \bibfield  {author} {\bibinfo {author} {\bibfnamefont {B.}~\bibnamefont
  {Abelev}} \emph {et~al.} (\bibinfo {collaboration} {ALICE}),\ }\href
  {\doibase 10.1007/JHEP03(2014)013} {\bibfield  {journal} {\bibinfo  {journal}
  {JHEP}\ }\textbf {\bibinfo {volume} {03}},\ \bibinfo {pages} {013} (\bibinfo
  {year} {2014})},\ \Eprint {http://arxiv.org/abs/1311.0633} {arXiv:1311.0633
  [nucl-ex]} \BibitemShut {NoStop}%
\bibitem [{\citenamefont {Adam}\ \emph {et~al.}(2015)\citenamefont {Adam} \emph
  {et~al.}}]{Adam:2015ewa}%
  \BibitemOpen
  \bibfield  {author} {\bibinfo {author} {\bibfnamefont {J.}~\bibnamefont
  {Adam}} \emph {et~al.} (\bibinfo {collaboration} {ALICE}),\ }\href {\doibase
  10.1016/j.physletb.2015.04.039} {\bibfield  {journal} {\bibinfo  {journal}
  {Phys. Lett. B}\ }\textbf {\bibinfo {volume} {746}},\ \bibinfo {pages} {1}
  (\bibinfo {year} {2015})},\ \Eprint {http://arxiv.org/abs/1502.01689}
  {arXiv:1502.01689 [nucl-ex]} \BibitemShut {NoStop}%
\bibitem [{\citenamefont {Aad}\ \emph {et~al.}(2015)\citenamefont {Aad},
  \citenamefont {Abbott}, \citenamefont {Abdallah}, \citenamefont {Khalek},
  \citenamefont {Abdinov}, \citenamefont {Aben}, \citenamefont {Abi},
  \citenamefont {Abolins}, \citenamefont {AbouZeid}, \citenamefont {Abramowicz}
  \emph {et~al.}}]{aad2015measurements}%
  \BibitemOpen
  \bibfield  {author} {\bibinfo {author} {\bibfnamefont {G.}~\bibnamefont
  {Aad}}, \bibinfo {author} {\bibfnamefont {B.}~\bibnamefont {Abbott}},
  \bibinfo {author} {\bibfnamefont {J.}~\bibnamefont {Abdallah}}, \bibinfo
  {author} {\bibfnamefont {S.~A.}\ \bibnamefont {Khalek}}, \bibinfo {author}
  {\bibfnamefont {O.}~\bibnamefont {Abdinov}}, \bibinfo {author} {\bibfnamefont
  {R.}~\bibnamefont {Aben}}, \bibinfo {author} {\bibfnamefont {B.}~\bibnamefont
  {Abi}}, \bibinfo {author} {\bibfnamefont {M.}~\bibnamefont {Abolins}},
  \bibinfo {author} {\bibfnamefont {O.}~\bibnamefont {AbouZeid}}, \bibinfo
  {author} {\bibfnamefont {H.}~\bibnamefont {Abramowicz}},  \emph {et~al.},\
  }\href@noop {} {\bibfield  {journal} {\bibinfo  {journal} {Phys. Rev. Lett.}\
  }\textbf {\bibinfo {volume} {114}},\ \bibinfo {pages} {072302} (\bibinfo
  {year} {2015})}\BibitemShut {NoStop}%
\bibitem [{\citenamefont {Aaboud}\ \emph {et~al.}(2017)\citenamefont {Aaboud}
  \emph {et~al.}}]{Aaboud:2017eww}%
  \BibitemOpen
  \bibfield  {author} {\bibinfo {author} {\bibfnamefont {M.}~\bibnamefont
  {Aaboud}} \emph {et~al.} (\bibinfo {collaboration} {ATLAS}),\ }\href
  {\doibase 10.1016/j.physletb.2017.09.078} {\bibfield  {journal} {\bibinfo
  {journal} {Phys. Lett. B}\ }\textbf {\bibinfo {volume} {774}},\ \bibinfo
  {pages} {379} (\bibinfo {year} {2017})},\ \Eprint
  {http://arxiv.org/abs/1706.09363} {arXiv:1706.09363 [hep-ex]} \BibitemShut
  {NoStop}%
\bibitem [{\citenamefont {Aaboud}\ \emph {et~al.}(2019)\citenamefont {Aaboud},
  \citenamefont {Aad}, \citenamefont {Abbott}, \citenamefont {Abdinov},
  \citenamefont {Abeloos}, \citenamefont {Abhayasinghe}, \citenamefont {Abidi},
  \citenamefont {AbouZeid}, \citenamefont {Abraham}, \citenamefont {Abramowicz}
  \emph {et~al.}}]{aaboud2019measurement}%
  \BibitemOpen
  \bibfield  {author} {\bibinfo {author} {\bibfnamefont {M.}~\bibnamefont
  {Aaboud}}, \bibinfo {author} {\bibfnamefont {G.}~\bibnamefont {Aad}},
  \bibinfo {author} {\bibfnamefont {B.}~\bibnamefont {Abbott}}, \bibinfo
  {author} {\bibfnamefont {O.}~\bibnamefont {Abdinov}}, \bibinfo {author}
  {\bibfnamefont {B.}~\bibnamefont {Abeloos}}, \bibinfo {author} {\bibfnamefont
  {D.~K.}\ \bibnamefont {Abhayasinghe}}, \bibinfo {author} {\bibfnamefont
  {S.~H.}\ \bibnamefont {Abidi}}, \bibinfo {author} {\bibfnamefont
  {O.}~\bibnamefont {AbouZeid}}, \bibinfo {author} {\bibfnamefont
  {N.}~\bibnamefont {Abraham}}, \bibinfo {author} {\bibfnamefont
  {H.}~\bibnamefont {Abramowicz}},  \emph {et~al.},\ }\href@noop {} {\bibfield
  {journal} {\bibinfo  {journal} {Phys. Lett. B}\ }\textbf {\bibinfo {volume}
  {790}},\ \bibinfo {pages} {108} (\bibinfo {year} {2019})}\BibitemShut
  {NoStop}%
\bibitem [{\citenamefont {Acharya}\ \emph {et~al.}(2020)\citenamefont {Acharya}
  \emph {et~al.}}]{Acharya:2019jyg}%
  \BibitemOpen
  \bibfield  {author} {\bibinfo {author} {\bibfnamefont {S.}~\bibnamefont
  {Acharya}} \emph {et~al.} (\bibinfo {collaboration} {ALICE}),\ }\href
  {\doibase 10.1103/PhysRevC.101.034911} {\bibfield  {journal} {\bibinfo
  {journal} {Phys. Rev. C}\ }\textbf {\bibinfo {volume} {101}},\ \bibinfo
  {pages} {034911} (\bibinfo {year} {2020})},\ \Eprint
  {http://arxiv.org/abs/1909.09718} {arXiv:1909.09718 [nucl-ex]} \BibitemShut
  {NoStop}%
\bibitem [{\citenamefont {Wang}\ and\ \citenamefont
  {Wang}(2002)}]{Wang:2002ri}%
  \BibitemOpen
  \bibfield  {author} {\bibinfo {author} {\bibfnamefont {E.}~\bibnamefont
  {Wang}}\ and\ \bibinfo {author} {\bibfnamefont {X.-N.}\ \bibnamefont
  {Wang}},\ }\href {\doibase 10.1103/PhysRevLett.89.162301} {\bibfield
  {journal} {\bibinfo  {journal} {Phys. Rev. Lett.}\ }\textbf {\bibinfo
  {volume} {89}},\ \bibinfo {pages} {162301} (\bibinfo {year} {2002})},\
  \Eprint {http://arxiv.org/abs/hep-ph/0202105} {arXiv:hep-ph/0202105}
  \BibitemShut {NoStop}%
\bibitem [{\citenamefont {Renk}(2006)}]{Renk:2006qg}%
  \BibitemOpen
  \bibfield  {author} {\bibinfo {author} {\bibfnamefont {T.}~\bibnamefont
  {Renk}},\ }\href {\doibase 10.1103/PhysRevC.74.034906} {\bibfield  {journal}
  {\bibinfo  {journal} {Phys. Rev. C}\ }\textbf {\bibinfo {volume} {74}},\
  \bibinfo {pages} {034906} (\bibinfo {year} {2006})},\ \Eprint
  {http://arxiv.org/abs/hep-ph/0607166} {arXiv:hep-ph/0607166} \BibitemShut
  {NoStop}%
\bibitem [{\citenamefont {Zhang}\ \emph {et~al.}(2007)\citenamefont {Zhang},
  \citenamefont {Owens}, \citenamefont {Wang},\ and\ \citenamefont
  {Wang}}]{Zhang:2007ja}%
  \BibitemOpen
  \bibfield  {author} {\bibinfo {author} {\bibfnamefont {H.}~\bibnamefont
  {Zhang}}, \bibinfo {author} {\bibfnamefont {J.~F.}\ \bibnamefont {Owens}},
  \bibinfo {author} {\bibfnamefont {E.}~\bibnamefont {Wang}}, \ and\ \bibinfo
  {author} {\bibfnamefont {X.-N.}\ \bibnamefont {Wang}},\ }\href {\doibase
  10.1103/PhysRevLett.98.212301} {\bibfield  {journal} {\bibinfo  {journal}
  {Phys. Rev. Lett.}\ }\textbf {\bibinfo {volume} {98}},\ \bibinfo {pages}
  {212301} (\bibinfo {year} {2007})},\ \Eprint
  {http://arxiv.org/abs/nucl-th/0701045} {arXiv:nucl-th/0701045} \BibitemShut
  {NoStop}%
\bibitem [{\citenamefont {Zhang}\ \emph {et~al.}(2009)\citenamefont {Zhang},
  \citenamefont {Owens}, \citenamefont {Wang},\ and\ \citenamefont
  {Wang}}]{Zhang:2009rn}%
  \BibitemOpen
  \bibfield  {author} {\bibinfo {author} {\bibfnamefont {H.}~\bibnamefont
  {Zhang}}, \bibinfo {author} {\bibfnamefont {J.~F.}\ \bibnamefont {Owens}},
  \bibinfo {author} {\bibfnamefont {E.}~\bibnamefont {Wang}}, \ and\ \bibinfo
  {author} {\bibfnamefont {X.-N.}\ \bibnamefont {Wang}},\ }\href {\doibase
  10.1103/PhysRevLett.103.032302} {\bibfield  {journal} {\bibinfo  {journal}
  {Phys. Rev. Lett.}\ }\textbf {\bibinfo {volume} {103}},\ \bibinfo {pages}
  {032302} (\bibinfo {year} {2009})},\ \Eprint {http://arxiv.org/abs/0902.4000}
  {arXiv:0902.4000 [nucl-th]} \BibitemShut {NoStop}%
\bibitem [{\citenamefont {He}\ \emph {et~al.}(2020)\citenamefont {He},
  \citenamefont {Pang},\ and\ \citenamefont {Wang}}]{He:2020iow}%
  \BibitemOpen
  \bibfield  {author} {\bibinfo {author} {\bibfnamefont {Y.}~\bibnamefont
  {He}}, \bibinfo {author} {\bibfnamefont {L.-G.}\ \bibnamefont {Pang}}, \ and\
  \bibinfo {author} {\bibfnamefont {X.-N.}\ \bibnamefont {Wang}},\ }\href
  {\doibase 10.1103/PhysRevLett.125.122301} {\bibfield  {journal} {\bibinfo
  {journal} {Phys. Rev. Lett.}\ }\textbf {\bibinfo {volume} {125}},\ \bibinfo
  {pages} {122301} (\bibinfo {year} {2020})},\ \Eprint
  {http://arxiv.org/abs/2001.08273} {arXiv:2001.08273 [hep-ph]} \BibitemShut
  {NoStop}%
\bibitem [{\citenamefont {Chen}\ \emph {et~al.}(2021)\citenamefont {Chen},
  \citenamefont {Yang}, \citenamefont {He}, \citenamefont {Ke}, \citenamefont
  {Pang},\ and\ \citenamefont {Wang}}]{PhysRevLett.127.082301}%
  \BibitemOpen
  \bibfield  {author} {\bibinfo {author} {\bibfnamefont {W.}~\bibnamefont
  {Chen}}, \bibinfo {author} {\bibfnamefont {Z.}~\bibnamefont {Yang}}, \bibinfo
  {author} {\bibfnamefont {Y.}~\bibnamefont {He}}, \bibinfo {author}
  {\bibfnamefont {W.}~\bibnamefont {Ke}}, \bibinfo {author} {\bibfnamefont
  {L.-G.}\ \bibnamefont {Pang}}, \ and\ \bibinfo {author} {\bibfnamefont
  {X.-N.}\ \bibnamefont {Wang}},\ }\href {\doibase
  10.1103/PhysRevLett.127.082301} {\bibfield  {journal} {\bibinfo  {journal}
  {Phys. Rev. Lett.}\ }\textbf {\bibinfo {volume} {127}},\ \bibinfo {pages}
  {082301} (\bibinfo {year} {2021})}\BibitemShut {NoStop}%
\bibitem [{\citenamefont {Armesto}\ \emph {et~al.}(2005)\citenamefont
  {Armesto}, \citenamefont {Salgado},\ and\ \citenamefont
  {Wiedemann}}]{Armesto:2004vz}%
  \BibitemOpen
  \bibfield  {author} {\bibinfo {author} {\bibfnamefont {N.}~\bibnamefont
  {Armesto}}, \bibinfo {author} {\bibfnamefont {C.~A.}\ \bibnamefont
  {Salgado}}, \ and\ \bibinfo {author} {\bibfnamefont {U.~A.}\ \bibnamefont
  {Wiedemann}},\ }\href {\doibase 10.1103/PhysRevC.72.064910} {\bibfield
  {journal} {\bibinfo  {journal} {Phys. Rev. C}\ }\textbf {\bibinfo {volume}
  {72}},\ \bibinfo {pages} {064910} (\bibinfo {year} {2005})},\ \Eprint
  {http://arxiv.org/abs/hep-ph/0411341} {arXiv:hep-ph/0411341} \BibitemShut
  {NoStop}%
\bibitem [{\citenamefont {Sadofyev}\ \emph {et~al.}(2021)\citenamefont
  {Sadofyev}, \citenamefont {Sievert},\ and\ \citenamefont
  {Vitev}}]{Sadofyev:2021ohn}%
  \BibitemOpen
  \bibfield  {author} {\bibinfo {author} {\bibfnamefont {A.~V.}\ \bibnamefont
  {Sadofyev}}, \bibinfo {author} {\bibfnamefont {M.~D.}\ \bibnamefont
  {Sievert}}, \ and\ \bibinfo {author} {\bibfnamefont {I.}~\bibnamefont
  {Vitev}},\ }\href@noop {} {\  (\bibinfo {year} {2021})},\ \Eprint
  {http://arxiv.org/abs/2104.09513} {arXiv:2104.09513 [hep-ph]} \BibitemShut
  {NoStop}%
\bibitem [{\citenamefont {Betz}\ and\ \citenamefont
  {Gyulassy}(2014)}]{Betz:2014cza}%
  \BibitemOpen
  \bibfield  {author} {\bibinfo {author} {\bibfnamefont {B.}~\bibnamefont
  {Betz}}\ and\ \bibinfo {author} {\bibfnamefont {M.}~\bibnamefont
  {Gyulassy}},\ }\href {\doibase 10.1007/JHEP10(2014)043} {\bibfield  {journal}
  {\bibinfo  {journal} {JHEP}\ }\textbf {\bibinfo {volume} {08}},\ \bibinfo
  {pages} {090} (\bibinfo {year} {2014})},\ \bibinfo {note} {[Erratum: JHEP 10,
  043 (2014)]},\ \Eprint {http://arxiv.org/abs/1404.6378} {arXiv:1404.6378
  [hep-ph]} \BibitemShut {NoStop}%
\bibitem [{\citenamefont {D'Eramo}\ \emph {et~al.}(2019)\citenamefont
  {D'Eramo}, \citenamefont {Rajagopal},\ and\ \citenamefont
  {Yin}}]{DEramo:2018eoy}%
  \BibitemOpen
  \bibfield  {author} {\bibinfo {author} {\bibfnamefont {F.}~\bibnamefont
  {D'Eramo}}, \bibinfo {author} {\bibfnamefont {K.}~\bibnamefont {Rajagopal}},
  \ and\ \bibinfo {author} {\bibfnamefont {Y.}~\bibnamefont {Yin}},\ }\href
  {\doibase 10.1007/JHEP01(2019)172} {\bibfield  {journal} {\bibinfo  {journal}
  {JHEP}\ }\textbf {\bibinfo {volume} {01}},\ \bibinfo {pages} {172} (\bibinfo
  {year} {2019})},\ \Eprint {http://arxiv.org/abs/1808.03250} {arXiv:1808.03250
  [hep-ph]} \BibitemShut {NoStop}%
\bibitem [{\citenamefont {Barata}\ \emph {et~al.}(2020)\citenamefont {Barata},
  \citenamefont {Mehtar-Tani}, \citenamefont {Soto-Ontoso},\ and\ \citenamefont
  {Tywoniuk}}]{Barata:2020rdn}%
  \BibitemOpen
  \bibfield  {author} {\bibinfo {author} {\bibfnamefont {J.~a.}\ \bibnamefont
  {Barata}}, \bibinfo {author} {\bibfnamefont {Y.}~\bibnamefont {Mehtar-Tani}},
  \bibinfo {author} {\bibfnamefont {A.}~\bibnamefont {Soto-Ontoso}}, \ and\
  \bibinfo {author} {\bibfnamefont {K.}~\bibnamefont {Tywoniuk}},\ }\href@noop
  {} {\  (\bibinfo {year} {2020})},\ \Eprint {http://arxiv.org/abs/2009.13667}
  {arXiv:2009.13667 [hep-ph]} \BibitemShut {NoStop}%
\bibitem [{\citenamefont {C\`e}\ \emph {et~al.}(2021)\citenamefont {C\`e},
  \citenamefont {Harris}, \citenamefont {Meyer},\ and\ \citenamefont
  {Toniato}}]{Harris:2020ijy}%
  \BibitemOpen
  \bibfield  {author} {\bibinfo {author} {\bibfnamefont {M.}~\bibnamefont
  {C\`e}}, \bibinfo {author} {\bibfnamefont {T.}~\bibnamefont {Harris}},
  \bibinfo {author} {\bibfnamefont {H.~B.}\ \bibnamefont {Meyer}}, \ and\
  \bibinfo {author} {\bibfnamefont {A.}~\bibnamefont {Toniato}},\ }\href
  {\doibase 10.1007/JHEP03(2021)035} {\bibfield  {journal} {\bibinfo  {journal}
  {JHEP}\ }\textbf {\bibinfo {volume} {03}},\ \bibinfo {pages} {035} (\bibinfo
  {year} {2021})},\ \Eprint {http://arxiv.org/abs/2012.07522} {arXiv:2012.07522
  [hep-ph]} \BibitemShut {NoStop}%
\bibitem [{\citenamefont {Baier}\ \emph {et~al.}(2001)\citenamefont {Baier},
  \citenamefont {Dokshitzer}, \citenamefont {Mueller},\ and\ \citenamefont
  {Schiff}}]{Baier:2001yt}%
  \BibitemOpen
  \bibfield  {author} {\bibinfo {author} {\bibfnamefont {R.}~\bibnamefont
  {Baier}}, \bibinfo {author} {\bibfnamefont {Y.~L.}\ \bibnamefont
  {Dokshitzer}}, \bibinfo {author} {\bibfnamefont {A.~H.}\ \bibnamefont
  {Mueller}}, \ and\ \bibinfo {author} {\bibfnamefont {D.}~\bibnamefont
  {Schiff}},\ }\href {\doibase 10.1088/1126-6708/2001/09/033} {\bibfield
  {journal} {\bibinfo  {journal} {JHEP}\ }\textbf {\bibinfo {volume} {09}},\
  \bibinfo {pages} {033} (\bibinfo {year} {2001})},\ \Eprint
  {http://arxiv.org/abs/hep-ph/0106347} {arXiv:hep-ph/0106347} \BibitemShut
  {NoStop}%
\bibitem [{\citenamefont {Du}\ \emph {et~al.}(2020)\citenamefont {Du},
  \citenamefont {Pablos},\ and\ \citenamefont {Tywoniuk}}]{Du:2020pmp}%
  \BibitemOpen
  \bibfield  {author} {\bibinfo {author} {\bibfnamefont {Y.-L.}\ \bibnamefont
  {Du}}, \bibinfo {author} {\bibfnamefont {D.}~\bibnamefont {Pablos}}, \ and\
  \bibinfo {author} {\bibfnamefont {K.}~\bibnamefont {Tywoniuk}},\ }\href
  {\doibase 10.1007/JHEP03(2021)206} {\bibfield  {journal} {\bibinfo  {journal}
  {JHEP}\ }\textbf {\bibinfo {volume} {21}},\ \bibinfo {pages} {206} (\bibinfo
  {year} {2020})},\ \Eprint {http://arxiv.org/abs/2012.07797} {arXiv:2012.07797
  [hep-ph]} \BibitemShut {NoStop}%
\bibitem [{\citenamefont {Dainese}\ \emph {et~al.}(2005)\citenamefont
  {Dainese}, \citenamefont {Loizides},\ and\ \citenamefont
  {Paic}}]{Dainese:2004te}%
  \BibitemOpen
  \bibfield  {author} {\bibinfo {author} {\bibfnamefont {A.}~\bibnamefont
  {Dainese}}, \bibinfo {author} {\bibfnamefont {C.}~\bibnamefont {Loizides}}, \
  and\ \bibinfo {author} {\bibfnamefont {G.}~\bibnamefont {Paic}},\ }\href
  {\doibase 10.1140/epjc/s2004-02077-x} {\bibfield  {journal} {\bibinfo
  {journal} {Eur. Phys. J. C}\ }\textbf {\bibinfo {volume} {38}},\ \bibinfo
  {pages} {461} (\bibinfo {year} {2005})},\ \Eprint
  {http://arxiv.org/abs/hep-ph/0406201} {arXiv:hep-ph/0406201} \BibitemShut
  {NoStop}%
\bibitem [{\citenamefont {Dasgupta}\ \emph {et~al.}(2015)\citenamefont
  {Dasgupta}, \citenamefont {Dreyer}, \citenamefont {Salam},\ and\
  \citenamefont {Soyez}}]{Dasgupta:2014yra}%
  \BibitemOpen
  \bibfield  {author} {\bibinfo {author} {\bibfnamefont {M.}~\bibnamefont
  {Dasgupta}}, \bibinfo {author} {\bibfnamefont {F.}~\bibnamefont {Dreyer}},
  \bibinfo {author} {\bibfnamefont {G.~P.}\ \bibnamefont {Salam}}, \ and\
  \bibinfo {author} {\bibfnamefont {G.}~\bibnamefont {Soyez}},\ }\href
  {\doibase 10.1007/JHEP04(2015)039} {\bibfield  {journal} {\bibinfo  {journal}
  {JHEP}\ }\textbf {\bibinfo {volume} {04}},\ \bibinfo {pages} {039} (\bibinfo
  {year} {2015})},\ \Eprint {http://arxiv.org/abs/1411.5182} {arXiv:1411.5182
  [hep-ph]} \BibitemShut {NoStop}%
\bibitem [{\citenamefont {Dasgupta}\ \emph {et~al.}(2016)\citenamefont
  {Dasgupta}, \citenamefont {Dreyer}, \citenamefont {Salam},\ and\
  \citenamefont {Soyez}}]{Dasgupta:2016bnd}%
  \BibitemOpen
  \bibfield  {author} {\bibinfo {author} {\bibfnamefont {M.}~\bibnamefont
  {Dasgupta}}, \bibinfo {author} {\bibfnamefont {F.~A.}\ \bibnamefont
  {Dreyer}}, \bibinfo {author} {\bibfnamefont {G.~P.}\ \bibnamefont {Salam}}, \
  and\ \bibinfo {author} {\bibfnamefont {G.}~\bibnamefont {Soyez}},\ }\href
  {\doibase 10.1007/JHEP06(2016)057} {\bibfield  {journal} {\bibinfo  {journal}
  {JHEP}\ }\textbf {\bibinfo {volume} {06}},\ \bibinfo {pages} {057} (\bibinfo
  {year} {2016})},\ \Eprint {http://arxiv.org/abs/1602.01110} {arXiv:1602.01110
  [hep-ph]} \BibitemShut {NoStop}%
\bibitem [{\citenamefont {Kang}\ \emph {et~al.}(2016)\citenamefont {Kang},
  \citenamefont {Ringer},\ and\ \citenamefont {Vitev}}]{Kang:2016mcy}%
  \BibitemOpen
  \bibfield  {author} {\bibinfo {author} {\bibfnamefont {Z.-B.}\ \bibnamefont
  {Kang}}, \bibinfo {author} {\bibfnamefont {F.}~\bibnamefont {Ringer}}, \ and\
  \bibinfo {author} {\bibfnamefont {I.}~\bibnamefont {Vitev}},\ }\href
  {\doibase 10.1007/JHEP10(2016)125} {\bibfield  {journal} {\bibinfo  {journal}
  {JHEP}\ }\textbf {\bibinfo {volume} {10}},\ \bibinfo {pages} {125} (\bibinfo
  {year} {2016})},\ \Eprint {http://arxiv.org/abs/1606.06732} {arXiv:1606.06732
  [hep-ph]} \BibitemShut {NoStop}%
\bibitem [{\citenamefont {Alioli}\ \emph {et~al.}(2019)\citenamefont {Alioli}
  \emph {et~al.}}]{Buckley:2019kjt}%
  \BibitemOpen
  \bibfield  {author} {\bibinfo {author} {\bibfnamefont {S.}~\bibnamefont
  {Alioli}} \emph {et~al.},\ }\href@noop {} {\  (\bibinfo {year} {2019})},\
  \Eprint {http://arxiv.org/abs/1902.01674} {arXiv:1902.01674 [hep-ph]}
  \BibitemShut {NoStop}%
\bibitem [{\citenamefont {Mehtar-Tani}\ and\ \citenamefont
  {Tywoniuk}(2018)}]{Mehtar-Tani:2017web}%
  \BibitemOpen
  \bibfield  {author} {\bibinfo {author} {\bibfnamefont {Y.}~\bibnamefont
  {Mehtar-Tani}}\ and\ \bibinfo {author} {\bibfnamefont {K.}~\bibnamefont
  {Tywoniuk}},\ }\href {\doibase 10.1103/PhysRevD.98.051501} {\bibfield
  {journal} {\bibinfo  {journal} {Phys. Rev. D}\ }\textbf {\bibinfo {volume}
  {98}},\ \bibinfo {pages} {051501(R)} (\bibinfo {year} {2018})},\ \Eprint
  {http://arxiv.org/abs/1707.07361} {arXiv:1707.07361 [hep-ph]} \BibitemShut
  {NoStop}%
\bibitem [{\citenamefont {Caucal}\ \emph {et~al.}(2018)\citenamefont {Caucal},
  \citenamefont {Iancu}, \citenamefont {Mueller},\ and\ \citenamefont
  {Soyez}}]{Caucal:2018dla}%
  \BibitemOpen
  \bibfield  {author} {\bibinfo {author} {\bibfnamefont {P.}~\bibnamefont
  {Caucal}}, \bibinfo {author} {\bibfnamefont {E.}~\bibnamefont {Iancu}},
  \bibinfo {author} {\bibfnamefont {A.~H.}\ \bibnamefont {Mueller}}, \ and\
  \bibinfo {author} {\bibfnamefont {G.}~\bibnamefont {Soyez}},\ }\href
  {\doibase 10.1103/PhysRevLett.120.232001} {\bibfield  {journal} {\bibinfo
  {journal} {Phys. Rev. Lett.}\ }\textbf {\bibinfo {volume} {120}},\ \bibinfo
  {pages} {232001} (\bibinfo {year} {2018})},\ \Eprint
  {http://arxiv.org/abs/1801.09703} {arXiv:1801.09703 [hep-ph]} \BibitemShut
  {NoStop}%
\bibitem [{\citenamefont {Dom\'\i{}nguez}\ \emph {et~al.}(2020)\citenamefont
  {Dom\'\i{}nguez}, \citenamefont {Milhano}, \citenamefont {Salgado},
  \citenamefont {Tywoniuk},\ and\ \citenamefont {Vila}}]{Dominguez:2019ges}%
  \BibitemOpen
  \bibfield  {author} {\bibinfo {author} {\bibfnamefont {F.}~\bibnamefont
  {Dom\'\i{}nguez}}, \bibinfo {author} {\bibfnamefont {J.~G.}\ \bibnamefont
  {Milhano}}, \bibinfo {author} {\bibfnamefont {C.~A.}\ \bibnamefont
  {Salgado}}, \bibinfo {author} {\bibfnamefont {K.}~\bibnamefont {Tywoniuk}}, \
  and\ \bibinfo {author} {\bibfnamefont {V.}~\bibnamefont {Vila}},\ }\href
  {\doibase 10.1140/epjc/s10052-019-7563-0} {\bibfield  {journal} {\bibinfo
  {journal} {Eur. Phys. J. C}\ }\textbf {\bibinfo {volume} {80}},\ \bibinfo
  {pages} {11} (\bibinfo {year} {2020})},\ \Eprint
  {http://arxiv.org/abs/1907.03653} {arXiv:1907.03653 [hep-ph]} \BibitemShut
  {NoStop}%
\bibitem [{\citenamefont {Casalderrey-Solana}\ \emph
  {et~al.}(2019)\citenamefont {Casalderrey-Solana}, \citenamefont {Hulcher},
  \citenamefont {Milhano}, \citenamefont {Pablos},\ and\ \citenamefont
  {Rajagopal}}]{Casalderrey-Solana:2018wrw}%
  \BibitemOpen
  \bibfield  {author} {\bibinfo {author} {\bibfnamefont {J.}~\bibnamefont
  {Casalderrey-Solana}}, \bibinfo {author} {\bibfnamefont {Z.}~\bibnamefont
  {Hulcher}}, \bibinfo {author} {\bibfnamefont {G.}~\bibnamefont {Milhano}},
  \bibinfo {author} {\bibfnamefont {D.}~\bibnamefont {Pablos}}, \ and\ \bibinfo
  {author} {\bibfnamefont {K.}~\bibnamefont {Rajagopal}},\ }\href {\doibase
  10.1103/PhysRevC.99.051901} {\bibfield  {journal} {\bibinfo  {journal} {Phys.
  Rev. C}\ }\textbf {\bibinfo {volume} {99}},\ \bibinfo {pages} {051901(R)}
  (\bibinfo {year} {2019})},\ \Eprint {http://arxiv.org/abs/1808.07386}
  {arXiv:1808.07386 [hep-ph]} \BibitemShut {NoStop}%
\bibitem [{\citenamefont {Casalderrey-Solana}\ \emph
  {et~al.}(2020)\citenamefont {Casalderrey-Solana}, \citenamefont {Milhano},
  \citenamefont {Pablos},\ and\ \citenamefont
  {Rajagopal}}]{Casalderrey-Solana:2019ubu}%
  \BibitemOpen
  \bibfield  {author} {\bibinfo {author} {\bibfnamefont {J.}~\bibnamefont
  {Casalderrey-Solana}}, \bibinfo {author} {\bibfnamefont {G.}~\bibnamefont
  {Milhano}}, \bibinfo {author} {\bibfnamefont {D.}~\bibnamefont {Pablos}}, \
  and\ \bibinfo {author} {\bibfnamefont {K.}~\bibnamefont {Rajagopal}},\ }\href
  {\doibase 10.1007/JHEP01(2020)044} {\bibfield  {journal} {\bibinfo  {journal}
  {JHEP}\ }\textbf {\bibinfo {volume} {01}},\ \bibinfo {pages} {044} (\bibinfo
  {year} {2020})},\ \Eprint {http://arxiv.org/abs/1907.11248} {arXiv:1907.11248
  [hep-ph]} \BibitemShut {NoStop}%
\bibitem [{\citenamefont {Caucal}\ \emph {et~al.}(2019)\citenamefont {Caucal},
  \citenamefont {Iancu},\ and\ \citenamefont {Soyez}}]{Caucal:2019uvr}%
  \BibitemOpen
  \bibfield  {author} {\bibinfo {author} {\bibfnamefont {P.}~\bibnamefont
  {Caucal}}, \bibinfo {author} {\bibfnamefont {E.}~\bibnamefont {Iancu}}, \
  and\ \bibinfo {author} {\bibfnamefont {G.}~\bibnamefont {Soyez}},\ }\href
  {\doibase 10.1007/JHEP10(2019)273} {\bibfield  {journal} {\bibinfo  {journal}
  {JHEP}\ }\textbf {\bibinfo {volume} {10}},\ \bibinfo {pages} {273} (\bibinfo
  {year} {2019})},\ \Eprint {http://arxiv.org/abs/1907.04866} {arXiv:1907.04866
  [hep-ph]} \BibitemShut {NoStop}%
\bibitem [{\citenamefont {Casalderrey-Solana}\ \emph
  {et~al.}(2015)\citenamefont {Casalderrey-Solana}, \citenamefont {Gulhan},
  \citenamefont {Milhano}, \citenamefont {Pablos},\ and\ \citenamefont
  {Rajagopal}}]{casalderrey2015erratum}%
  \BibitemOpen
  \bibfield  {author} {\bibinfo {author} {\bibfnamefont {J.}~\bibnamefont
  {Casalderrey-Solana}}, \bibinfo {author} {\bibfnamefont {D.~C.}\ \bibnamefont
  {Gulhan}}, \bibinfo {author} {\bibfnamefont {J.~G.}\ \bibnamefont {Milhano}},
  \bibinfo {author} {\bibfnamefont {D.}~\bibnamefont {Pablos}}, \ and\ \bibinfo
  {author} {\bibfnamefont {K.}~\bibnamefont {Rajagopal}},\ }\href@noop {}
  {\bibfield  {journal} {\bibinfo  {journal} {JHEP}\ }\textbf {\bibinfo
  {volume} {2015}},\ \bibinfo {pages} {175} (\bibinfo {year}
  {2015})}\BibitemShut {NoStop}%
\bibitem [{\citenamefont {Casalderrey-Solana}\ \emph
  {et~al.}(2016)\citenamefont {Casalderrey-Solana}, \citenamefont {Gulhan},
  \citenamefont {Milhano}, \citenamefont {Pablos},\ and\ \citenamefont
  {Rajagopal}}]{Casalderrey-Solana:2015vaa}%
  \BibitemOpen
  \bibfield  {author} {\bibinfo {author} {\bibfnamefont {J.}~\bibnamefont
  {Casalderrey-Solana}}, \bibinfo {author} {\bibfnamefont {D.~C.}\ \bibnamefont
  {Gulhan}}, \bibinfo {author} {\bibfnamefont {J.~G.}\ \bibnamefont {Milhano}},
  \bibinfo {author} {\bibfnamefont {D.}~\bibnamefont {Pablos}}, \ and\ \bibinfo
  {author} {\bibfnamefont {K.}~\bibnamefont {Rajagopal}},\ }\href {\doibase
  10.1007/JHEP03(2016)053} {\bibfield  {journal} {\bibinfo  {journal} {JHEP}\
  }\textbf {\bibinfo {volume} {03}},\ \bibinfo {pages} {053} (\bibinfo {year}
  {2016})},\ \Eprint {http://arxiv.org/abs/1508.00815} {arXiv:1508.00815
  [hep-ph]} \BibitemShut {NoStop}%
\bibitem [{\citenamefont {Casalderrey-Solana}\ \emph
  {et~al.}(2017)\citenamefont {Casalderrey-Solana}, \citenamefont {Gulhan},
  \citenamefont {Milhano}, \citenamefont {Pablos},\ and\ \citenamefont
  {Rajagopal}}]{Casalderrey-Solana:2016jvj}%
  \BibitemOpen
  \bibfield  {author} {\bibinfo {author} {\bibfnamefont {J.}~\bibnamefont
  {Casalderrey-Solana}}, \bibinfo {author} {\bibfnamefont {D.}~\bibnamefont
  {Gulhan}}, \bibinfo {author} {\bibfnamefont {G.}~\bibnamefont {Milhano}},
  \bibinfo {author} {\bibfnamefont {D.}~\bibnamefont {Pablos}}, \ and\ \bibinfo
  {author} {\bibfnamefont {K.}~\bibnamefont {Rajagopal}},\ }\href {\doibase
  10.1007/JHEP03(2017)135} {\bibfield  {journal} {\bibinfo  {journal} {JHEP}\
  }\textbf {\bibinfo {volume} {03}},\ \bibinfo {pages} {135} (\bibinfo {year}
  {2017})},\ \Eprint {http://arxiv.org/abs/1609.05842} {arXiv:1609.05842
  [hep-ph]} \BibitemShut {NoStop}%
\bibitem [{Note1()}]{Note1}%
  \BibitemOpen
  \bibinfo {note} {\protect \leavevmode {\protect \color {black}See the
  supplemental material for a preliminary check on the model dependence of the
  extraction of $\chi $.}}\BibitemShut {Stop}%
\bibitem [{\citenamefont {Apolin\'ario}\ \emph {et~al.}(2021)\citenamefont
  {Apolin\'ario}, \citenamefont {Castro}, \citenamefont {Crispim Rom\~ao},
  \citenamefont {Milhano}, \citenamefont {Pedro},\ and\ \citenamefont
  {Peres}}]{Apolinario:2021olp}%
  \BibitemOpen
  \bibfield  {author} {\bibinfo {author} {\bibfnamefont {L.}~\bibnamefont
  {Apolin\'ario}}, \bibinfo {author} {\bibfnamefont {N.~F.}\ \bibnamefont
  {Castro}}, \bibinfo {author} {\bibfnamefont {M.}~\bibnamefont {Crispim
  Rom\~ao}}, \bibinfo {author} {\bibfnamefont {J.~G.}\ \bibnamefont {Milhano}},
  \bibinfo {author} {\bibfnamefont {R.}~\bibnamefont {Pedro}}, \ and\ \bibinfo
  {author} {\bibfnamefont {F.~C.~R.}\ \bibnamefont {Peres}},\ }\href@noop {} {\
   (\bibinfo {year} {2021})},\ \Eprint {http://arxiv.org/abs/2106.08869}
  {arXiv:2106.08869 [hep-ph]} \BibitemShut {NoStop}%
\bibitem [{\citenamefont {Bia{\l}as}\ \emph {et~al.}(1976)\citenamefont
  {Bia{\l}as}, \citenamefont {Bleszy{\'n}ski},\ and\ \citenamefont
  {Czy{\.z}}}]{biallas1976multiplicity}%
  \BibitemOpen
  \bibfield  {author} {\bibinfo {author} {\bibfnamefont {A.}~\bibnamefont
  {Bia{\l}as}}, \bibinfo {author} {\bibfnamefont {M.}~\bibnamefont
  {Bleszy{\'n}ski}}, \ and\ \bibinfo {author} {\bibfnamefont {W.}~\bibnamefont
  {Czy{\.z}}},\ }\href@noop {} {\bibfield  {journal} {\bibinfo  {journal}
  {Nucl. Phys. B}\ }\textbf {\bibinfo {volume} {111}},\ \bibinfo {pages} {461}
  (\bibinfo {year} {1976})}\BibitemShut {NoStop}%
\bibitem [{\citenamefont {Miller}\ \emph {et~al.}(2007)\citenamefont {Miller},
  \citenamefont {Reygers}, \citenamefont {Sanders},\ and\ \citenamefont
  {Steinberg}}]{miller2007glauber}%
  \BibitemOpen
  \bibfield  {author} {\bibinfo {author} {\bibfnamefont {M.~L.}\ \bibnamefont
  {Miller}}, \bibinfo {author} {\bibfnamefont {K.}~\bibnamefont {Reygers}},
  \bibinfo {author} {\bibfnamefont {S.~J.}\ \bibnamefont {Sanders}}, \ and\
  \bibinfo {author} {\bibfnamefont {P.}~\bibnamefont {Steinberg}},\ }\href@noop
  {} {\bibfield  {journal} {\bibinfo  {journal} {Annu. Rev. Nucl. Part. Sci.}\
  }\textbf {\bibinfo {volume} {57}},\ \bibinfo {pages} {205} (\bibinfo {year}
  {2007})}\BibitemShut {NoStop}%
\bibitem [{\citenamefont {Cacciari}\ \emph {et~al.}(2008)\citenamefont
  {Cacciari}, \citenamefont {Salam},\ and\ \citenamefont
  {Soyez}}]{Cacciari:2008gp}%
  \BibitemOpen
  \bibfield  {author} {\bibinfo {author} {\bibfnamefont {M.}~\bibnamefont
  {Cacciari}}, \bibinfo {author} {\bibfnamefont {G.~P.}\ \bibnamefont {Salam}},
  \ and\ \bibinfo {author} {\bibfnamefont {G.}~\bibnamefont {Soyez}},\ }\href
  {\doibase 10.1088/1126-6708/2008/04/063} {\bibfield  {journal} {\bibinfo
  {journal} {JHEP}\ }\textbf {\bibinfo {volume} {04}},\ \bibinfo {pages} {063}
  (\bibinfo {year} {2008})},\ \Eprint {http://arxiv.org/abs/0802.1189}
  {arXiv:0802.1189 [hep-ph]} \BibitemShut {NoStop}%
\bibitem [{\citenamefont {Cacciari}\ \emph {et~al.}(2012)\citenamefont
  {Cacciari}, \citenamefont {Salam},\ and\ \citenamefont
  {Soyez}}]{Cacciari:2011ma}%
  \BibitemOpen
  \bibfield  {author} {\bibinfo {author} {\bibfnamefont {M.}~\bibnamefont
  {Cacciari}}, \bibinfo {author} {\bibfnamefont {G.~P.}\ \bibnamefont {Salam}},
  \ and\ \bibinfo {author} {\bibfnamefont {G.}~\bibnamefont {Soyez}},\ }\href
  {\doibase 10.1140/epjc/s10052-012-1896-2} {\bibfield  {journal} {\bibinfo
  {journal} {Eur. Phys. J. C}\ }\textbf {\bibinfo {volume} {72}},\ \bibinfo
  {pages} {1896} (\bibinfo {year} {2012})},\ \Eprint
  {http://arxiv.org/abs/1111.6097} {arXiv:1111.6097 [hep-ph]} \BibitemShut
  {NoStop}%
\bibitem [{Note2()}]{Note2}%
  \BibitemOpen
  \bibinfo {note} {Additionally, we demand that $p_{\scriptscriptstyle T}>100$
  GeV because this corresponds to the kinematical range where we have trained
  our neural network. We have also checked that our results do not depend on
  the precise cut on $p_{\scriptscriptstyle T}$ as long as it is sufficiently
  below the cut of $p_{\scriptscriptstyle T}^{\protect \rm \relax \protect
  \fontsize {5}{6}\protect \selectfont initial}$.}\BibitemShut {Stop}%
\bibitem [{\citenamefont {Wiedemann}(2008)}]{Wiedemann:2008zz}%
  \BibitemOpen
  \bibfield  {author} {\bibinfo {author} {\bibfnamefont {U.~A.}\ \bibnamefont
  {Wiedemann}},\ }in\ \href@noop {} {\emph {\bibinfo {booktitle} {{2007
  European School of High-Energy Physics}}}}\ (\bibinfo {year}
  {2008})\BibitemShut {NoStop}%
\bibitem [{\citenamefont {Petersen}\ \emph {et~al.}(2012)\citenamefont
  {Petersen}, \citenamefont {La~Placa},\ and\ \citenamefont
  {Bass}}]{Petersen:2012qc}%
  \BibitemOpen
  \bibfield  {author} {\bibinfo {author} {\bibfnamefont {H.}~\bibnamefont
  {Petersen}}, \bibinfo {author} {\bibfnamefont {R.}~\bibnamefont {La~Placa}},
  \ and\ \bibinfo {author} {\bibfnamefont {S.~A.}\ \bibnamefont {Bass}},\
  }\href {\doibase 10.1088/0954-3899/39/5/055102} {\bibfield  {journal}
  {\bibinfo  {journal} {J. Phys. G}\ }\textbf {\bibinfo {volume} {39}},\
  \bibinfo {pages} {055102} (\bibinfo {year} {2012})},\ \Eprint
  {http://arxiv.org/abs/1201.1881} {arXiv:1201.1881 [nucl-th]} \BibitemShut
  {NoStop}%
\bibitem [{\citenamefont {Greif}\ \emph {et~al.}(2017)\citenamefont {Greif},
  \citenamefont {Greiner}, \citenamefont {Schenke}, \citenamefont
  {Schlichting},\ and\ \citenamefont {Xu}}]{Greif:2017bnr}%
  \BibitemOpen
  \bibfield  {author} {\bibinfo {author} {\bibfnamefont {M.}~\bibnamefont
  {Greif}}, \bibinfo {author} {\bibfnamefont {C.}~\bibnamefont {Greiner}},
  \bibinfo {author} {\bibfnamefont {B.}~\bibnamefont {Schenke}}, \bibinfo
  {author} {\bibfnamefont {S.}~\bibnamefont {Schlichting}}, \ and\ \bibinfo
  {author} {\bibfnamefont {Z.}~\bibnamefont {Xu}},\ }\href {\doibase
  10.1103/PhysRevD.96.091504} {\bibfield  {journal} {\bibinfo  {journal} {Phys.
  Rev. D}\ }\textbf {\bibinfo {volume} {96}},\ \bibinfo {pages} {091504(R)}
  (\bibinfo {year} {2017})},\ \Eprint {http://arxiv.org/abs/1708.02076}
  {arXiv:1708.02076 [hep-ph]} \BibitemShut {NoStop}%
\bibitem [{\citenamefont {Nie}\ \emph {et~al.}(2019)\citenamefont {Nie},
  \citenamefont {Yi}, \citenamefont {Luo}, \citenamefont {Ma},\ and\
  \citenamefont {Jia}}]{Nie:2019swk}%
  \BibitemOpen
  \bibfield  {author} {\bibinfo {author} {\bibfnamefont {M.}~\bibnamefont
  {Nie}}, \bibinfo {author} {\bibfnamefont {L.}~\bibnamefont {Yi}}, \bibinfo
  {author} {\bibfnamefont {X.}~\bibnamefont {Luo}}, \bibinfo {author}
  {\bibfnamefont {G.}~\bibnamefont {Ma}}, \ and\ \bibinfo {author}
  {\bibfnamefont {J.}~\bibnamefont {Jia}},\ }\href {\doibase
  10.1103/PhysRevC.100.064905} {\bibfield  {journal} {\bibinfo  {journal}
  {Phys. Rev. C}\ }\textbf {\bibinfo {volume} {100}},\ \bibinfo {pages}
  {064905} (\bibinfo {year} {2019})},\ \Eprint
  {http://arxiv.org/abs/1906.01422} {arXiv:1906.01422 [nucl-th]} \BibitemShut
  {NoStop}%
\bibitem [{\citenamefont {Giacalone}\ \emph {et~al.}(2020)\citenamefont
  {Giacalone}, \citenamefont {Schenke},\ and\ \citenamefont
  {Shen}}]{Giacalone:2020byk}%
  \BibitemOpen
  \bibfield  {author} {\bibinfo {author} {\bibfnamefont {G.}~\bibnamefont
  {Giacalone}}, \bibinfo {author} {\bibfnamefont {B.}~\bibnamefont {Schenke}},
  \ and\ \bibinfo {author} {\bibfnamefont {C.}~\bibnamefont {Shen}},\ }\href
  {\doibase 10.1103/PhysRevLett.125.192301} {\bibfield  {journal} {\bibinfo
  {journal} {Phys. Rev. Lett.}\ }\textbf {\bibinfo {volume} {125}},\ \bibinfo
  {pages} {192301} (\bibinfo {year} {2020})},\ \Eprint
  {http://arxiv.org/abs/2006.15721} {arXiv:2006.15721 [nucl-th]} \BibitemShut
  {NoStop}%
\bibitem [{\citenamefont {Gyulassy}\ \emph {et~al.}(2001)\citenamefont
  {Gyulassy}, \citenamefont {Vitev},\ and\ \citenamefont
  {Wang}}]{gyulassy2001high}%
  \BibitemOpen
  \bibfield  {author} {\bibinfo {author} {\bibfnamefont {M.}~\bibnamefont
  {Gyulassy}}, \bibinfo {author} {\bibfnamefont {I.}~\bibnamefont {Vitev}}, \
  and\ \bibinfo {author} {\bibfnamefont {X.-N.}\ \bibnamefont {Wang}},\
  }\href@noop {} {\bibfield  {journal} {\bibinfo  {journal} {Phys. Rev. Lett.}\
  }\textbf {\bibinfo {volume} {86}},\ \bibinfo {pages} {2537} (\bibinfo {year}
  {2001})}\BibitemShut {NoStop}%
\bibitem [{\citenamefont {Wang}(2001)}]{wang2001jet}%
  \BibitemOpen
  \bibfield  {author} {\bibinfo {author} {\bibfnamefont {X.-N.}\ \bibnamefont
  {Wang}},\ }\href@noop {} {\bibfield  {journal} {\bibinfo  {journal} {Phys.
  Rev. C}\ }\textbf {\bibinfo {volume} {63}},\ \bibinfo {pages} {054902}
  (\bibinfo {year} {2001})}\BibitemShut {NoStop}%
\bibitem [{\citenamefont {Aad}\ \emph {et~al.}(2013)\citenamefont {Aad},
  \citenamefont {Abajyan}, \citenamefont {Abbott}, \citenamefont {Abdallah},
  \citenamefont {Khalek}, \citenamefont {Aben}, \citenamefont {Abi},
  \citenamefont {Abolins}, \citenamefont {AbouZeid}, \citenamefont {Abramowicz}
  \emph {et~al.}}]{aad2013measurement}%
  \BibitemOpen
  \bibfield  {author} {\bibinfo {author} {\bibfnamefont {G.}~\bibnamefont
  {Aad}}, \bibinfo {author} {\bibfnamefont {T.}~\bibnamefont {Abajyan}},
  \bibinfo {author} {\bibfnamefont {B.}~\bibnamefont {Abbott}}, \bibinfo
  {author} {\bibfnamefont {J.}~\bibnamefont {Abdallah}}, \bibinfo {author}
  {\bibfnamefont {S.~A.}\ \bibnamefont {Khalek}}, \bibinfo {author}
  {\bibfnamefont {R.}~\bibnamefont {Aben}}, \bibinfo {author} {\bibfnamefont
  {B.}~\bibnamefont {Abi}}, \bibinfo {author} {\bibfnamefont {M.}~\bibnamefont
  {Abolins}}, \bibinfo {author} {\bibfnamefont {O.}~\bibnamefont {AbouZeid}},
  \bibinfo {author} {\bibfnamefont {H.}~\bibnamefont {Abramowicz}},  \emph
  {et~al.} (\bibinfo {collaboration} {ATLAS Collaboration}),\ }\href {\doibase
  10.1103/PhysRevLett.111.152301} {\bibfield  {journal} {\bibinfo  {journal}
  {Phys. Rev. Lett}\ }\textbf {\bibinfo {volume} {111}},\ \bibinfo {pages}
  {152301} (\bibinfo {year} {2013})}\BibitemShut {NoStop}%
\bibitem [{\citenamefont {Lappi}\ \emph {et~al.}(2016)\citenamefont {Lappi},
  \citenamefont {Schenke}, \citenamefont {Schlichting},\ and\ \citenamefont
  {Venugopalan}}]{lappi2016tracing}%
  \BibitemOpen
  \bibfield  {author} {\bibinfo {author} {\bibfnamefont {T.}~\bibnamefont
  {Lappi}}, \bibinfo {author} {\bibfnamefont {B.}~\bibnamefont {Schenke}},
  \bibinfo {author} {\bibfnamefont {S.}~\bibnamefont {Schlichting}}, \ and\
  \bibinfo {author} {\bibfnamefont {R.}~\bibnamefont {Venugopalan}},\
  }\href@noop {} {\bibfield  {journal} {\bibinfo  {journal} {JHEP}\ }\textbf
  {\bibinfo {volume} {2016}},\ \bibinfo {pages} {61} (\bibinfo {year}
  {2016})}\BibitemShut {NoStop}%
\bibitem [{\citenamefont {Mace}\ \emph {et~al.}(2018)\citenamefont {Mace},
  \citenamefont {Skokov}, \citenamefont {Tribedy},\ and\ \citenamefont
  {Venugopalan}}]{mace2018hierarchy}%
  \BibitemOpen
  \bibfield  {author} {\bibinfo {author} {\bibfnamefont {M.}~\bibnamefont
  {Mace}}, \bibinfo {author} {\bibfnamefont {V.~V.}\ \bibnamefont {Skokov}},
  \bibinfo {author} {\bibfnamefont {P.}~\bibnamefont {Tribedy}}, \ and\
  \bibinfo {author} {\bibfnamefont {R.}~\bibnamefont {Venugopalan}},\
  }\href@noop {} {\bibfield  {journal} {\bibinfo  {journal} {Phys. Rev. Lett.}\
  }\textbf {\bibinfo {volume} {121}},\ \bibinfo {pages} {052301} (\bibinfo
  {year} {2018})}\BibitemShut {NoStop}%
\bibitem [{\citenamefont {Zhang}\ \emph {et~al.}(2019)\citenamefont {Zhang},
  \citenamefont {Marquet}, \citenamefont {Qin}, \citenamefont {Wei},\ and\
  \citenamefont {Xiao}}]{PhysRevLett.122.172302}%
  \BibitemOpen
  \bibfield  {author} {\bibinfo {author} {\bibfnamefont {C.}~\bibnamefont
  {Zhang}}, \bibinfo {author} {\bibfnamefont {C.}~\bibnamefont {Marquet}},
  \bibinfo {author} {\bibfnamefont {G.-Y.}\ \bibnamefont {Qin}}, \bibinfo
  {author} {\bibfnamefont {S.-Y.}\ \bibnamefont {Wei}}, \ and\ \bibinfo
  {author} {\bibfnamefont {B.-W.}\ \bibnamefont {Xiao}},\ }\href {\doibase
  10.1103/PhysRevLett.122.172302} {\bibfield  {journal} {\bibinfo  {journal}
  {Phys. Rev. Lett.}\ }\textbf {\bibinfo {volume} {122}},\ \bibinfo {pages}
  {172302} (\bibinfo {year} {2019})}\BibitemShut {NoStop}%
\bibitem [{\citenamefont {Albacete}\ \emph {et~al.}(2017)\citenamefont
  {Albacete}, \citenamefont {Petersen},\ and\ \citenamefont
  {Soto-Ontoso}}]{Albacete:2016gxu}%
  \BibitemOpen
  \bibfield  {author} {\bibinfo {author} {\bibfnamefont {J.~L.}\ \bibnamefont
  {Albacete}}, \bibinfo {author} {\bibfnamefont {H.}~\bibnamefont {Petersen}},
  \ and\ \bibinfo {author} {\bibfnamefont {A.}~\bibnamefont {Soto-Ontoso}},\
  }\href {\doibase 10.1103/PhysRevC.95.064909} {\bibfield  {journal} {\bibinfo
  {journal} {Phys. Rev. C}\ }\textbf {\bibinfo {volume} {95}},\ \bibinfo
  {pages} {064909} (\bibinfo {year} {2017})},\ \Eprint
  {http://arxiv.org/abs/1612.06274} {arXiv:1612.06274 [hep-ph]} \BibitemShut
  {NoStop}%
\bibitem [{\citenamefont {Gelis}\ \emph {et~al.}(2019)\citenamefont {Gelis},
  \citenamefont {Giacalone}, \citenamefont {Guerrero-Rodr{\'\i}guez},
  \citenamefont {Marquet},\ and\ \citenamefont
  {Ollitrault}}]{gelis2019primordial}%
  \BibitemOpen
  \bibfield  {author} {\bibinfo {author} {\bibfnamefont {F.}~\bibnamefont
  {Gelis}}, \bibinfo {author} {\bibfnamefont {G.}~\bibnamefont {Giacalone}},
  \bibinfo {author} {\bibfnamefont {P.}~\bibnamefont
  {Guerrero-Rodr{\'\i}guez}}, \bibinfo {author} {\bibfnamefont
  {C.}~\bibnamefont {Marquet}}, \ and\ \bibinfo {author} {\bibfnamefont
  {J.-Y.}\ \bibnamefont {Ollitrault}},\ }\href@noop {} {\bibfield  {journal}
  {\bibinfo  {journal} {arXiv preprint arXiv:1907.10948}\ } (\bibinfo {year}
  {2019})}\BibitemShut {NoStop}%
\bibitem [{\citenamefont {Blok}\ \emph {et~al.}(2017)\citenamefont {Blok},
  \citenamefont {J\"akel}, \citenamefont {Strikman},\ and\ \citenamefont
  {Wiedemann}}]{Blok:2017pui}%
  \BibitemOpen
  \bibfield  {author} {\bibinfo {author} {\bibfnamefont {B.}~\bibnamefont
  {Blok}}, \bibinfo {author} {\bibfnamefont {C.~D.}\ \bibnamefont {J\"akel}},
  \bibinfo {author} {\bibfnamefont {M.}~\bibnamefont {Strikman}}, \ and\
  \bibinfo {author} {\bibfnamefont {U.~A.}\ \bibnamefont {Wiedemann}},\ }\href
  {\doibase 10.1007/JHEP12(2017)074} {\bibfield  {journal} {\bibinfo  {journal}
  {JHEP}\ }\textbf {\bibinfo {volume} {12}},\ \bibinfo {pages} {074} (\bibinfo
  {year} {2017})},\ \Eprint {http://arxiv.org/abs/1708.08241} {arXiv:1708.08241
  [hep-ph]} \BibitemShut {NoStop}%
\bibitem [{\citenamefont {Blok}\ and\ \citenamefont
  {Wiedemann}(2019)}]{Blok:2018xes}%
  \BibitemOpen
  \bibfield  {author} {\bibinfo {author} {\bibfnamefont {B.}~\bibnamefont
  {Blok}}\ and\ \bibinfo {author} {\bibfnamefont {U.~A.}\ \bibnamefont
  {Wiedemann}},\ }\href {\doibase 10.1016/j.physletb.2019.05.038} {\bibfield
  {journal} {\bibinfo  {journal} {Phys. Lett. B}\ }\textbf {\bibinfo {volume}
  {795}},\ \bibinfo {pages} {259} (\bibinfo {year} {2019})},\ \Eprint
  {http://arxiv.org/abs/1812.04113} {arXiv:1812.04113 [hep-ph]} \BibitemShut
  {NoStop}%
\bibitem [{Note3()}]{Note3}%
  \BibitemOpen
  \bibinfo {note} {In principle, given the absence of energy loss in colorless
  particles, the value of $v_2$ for high-$p_T$ photons \cite {Hamed:2014hta} or
  $Z^0$ bosons should also be sensitive to such initial-state induced
  anisotropy.}\BibitemShut {Stop}%
\bibitem [{\citenamefont {Milhano}\ and\ \citenamefont
  {Zapp}(2016)}]{Milhano:2015mng}%
  \BibitemOpen
  \bibfield  {author} {\bibinfo {author} {\bibfnamefont {J.~G.}\ \bibnamefont
  {Milhano}}\ and\ \bibinfo {author} {\bibfnamefont {K.~C.}\ \bibnamefont
  {Zapp}},\ }\href {\doibase 10.1140/epjc/s10052-016-4130-9} {\bibfield
  {journal} {\bibinfo  {journal} {Eur. Phys. J. C}\ }\textbf {\bibinfo {volume}
  {76}},\ \bibinfo {pages} {288} (\bibinfo {year} {2016})},\ \Eprint
  {http://arxiv.org/abs/1512.08107} {arXiv:1512.08107 [hep-ph]} \BibitemShut
  {NoStop}%
\bibitem [{Note4()}]{Note4}%
  \BibitemOpen
  \bibinfo {note} {A reasonable estimate of the amount of medium induced
  modification can also be obtained from the ratio between the jet $p_T$ and
  that of a recoiling colorless trigger boson, although the correlation is not
  tight even in vacuum due to sizeable out-of-cone radiation \cite
  {Zhang:2009rn}.}\BibitemShut {Stop}%
\bibitem [{\citenamefont {Brewer}\ \emph {et~al.}(2019)\citenamefont {Brewer},
  \citenamefont {Milhano},\ and\ \citenamefont {Thaler}}]{Brewer:2018dfs}%
  \BibitemOpen
  \bibfield  {author} {\bibinfo {author} {\bibfnamefont {J.}~\bibnamefont
  {Brewer}}, \bibinfo {author} {\bibfnamefont {J.~G.}\ \bibnamefont {Milhano}},
  \ and\ \bibinfo {author} {\bibfnamefont {J.}~\bibnamefont {Thaler}},\ }\href
  {\doibase 10.1103/PhysRevLett.122.222301} {\bibfield  {journal} {\bibinfo
  {journal} {Phys. Rev. Lett.}\ }\textbf {\bibinfo {volume} {122}},\ \bibinfo
  {pages} {222301} (\bibinfo {year} {2019})},\ \Eprint
  {http://arxiv.org/abs/1812.05111} {arXiv:1812.05111 [hep-ph]} \BibitemShut
  {NoStop}%
\bibitem [{\citenamefont {Takacs}\ and\ \citenamefont
  {Tywoniuk}(2021)}]{Takacs:2021bpv}%
  \BibitemOpen
  \bibfield  {author} {\bibinfo {author} {\bibfnamefont {A.}~\bibnamefont
  {Takacs}}\ and\ \bibinfo {author} {\bibfnamefont {K.}~\bibnamefont
  {Tywoniuk}},\ }\href@noop {} {\  (\bibinfo {year} {2021})},\ \Eprint
  {http://arxiv.org/abs/2103.14676} {arXiv:2103.14676 [hep-ph]} \BibitemShut
  {NoStop}%
\bibitem [{\citenamefont {Yan}\ \emph {et~al.}(2018)\citenamefont {Yan},
  \citenamefont {Jeon},\ and\ \citenamefont {Gale}}]{Yan:2017rku}%
  \BibitemOpen
  \bibfield  {author} {\bibinfo {author} {\bibfnamefont {L.}~\bibnamefont
  {Yan}}, \bibinfo {author} {\bibfnamefont {S.}~\bibnamefont {Jeon}}, \ and\
  \bibinfo {author} {\bibfnamefont {C.}~\bibnamefont {Gale}},\ }\href {\doibase
  10.1103/PhysRevC.97.034914} {\bibfield  {journal} {\bibinfo  {journal} {Phys.
  Rev. C}\ }\textbf {\bibinfo {volume} {97}},\ \bibinfo {pages} {034914}
  (\bibinfo {year} {2018})},\ \Eprint {http://arxiv.org/abs/1707.09519}
  {arXiv:1707.09519 [nucl-th]} \BibitemShut {NoStop}%
\bibitem [{\citenamefont {Tachibana}\ \emph {et~al.}(2020)\citenamefont
  {Tachibana}, \citenamefont {Shen},\ and\ \citenamefont
  {Majumder}}]{Tachibana:2020mtb}%
  \BibitemOpen
  \bibfield  {author} {\bibinfo {author} {\bibfnamefont {Y.}~\bibnamefont
  {Tachibana}}, \bibinfo {author} {\bibfnamefont {C.}~\bibnamefont {Shen}}, \
  and\ \bibinfo {author} {\bibfnamefont {A.}~\bibnamefont {Majumder}},\
  }\href@noop {} {\  (\bibinfo {year} {2020})},\ \Eprint
  {http://arxiv.org/abs/2001.08321} {arXiv:2001.08321 [nucl-th]} \BibitemShut
  {NoStop}%
\bibitem [{\citenamefont {Casalderrey-Solana}\ \emph
  {et~al.}(2021)\citenamefont {Casalderrey-Solana}, \citenamefont {Milhano},
  \citenamefont {Pablos}, \citenamefont {Rajagopal},\ and\ \citenamefont
  {Yao}}]{Casalderrey-Solana:2020rsj}%
  \BibitemOpen
  \bibfield  {author} {\bibinfo {author} {\bibfnamefont {J.}~\bibnamefont
  {Casalderrey-Solana}}, \bibinfo {author} {\bibfnamefont {J.~G.}\ \bibnamefont
  {Milhano}}, \bibinfo {author} {\bibfnamefont {D.}~\bibnamefont {Pablos}},
  \bibinfo {author} {\bibfnamefont {K.}~\bibnamefont {Rajagopal}}, \ and\
  \bibinfo {author} {\bibfnamefont {X.}~\bibnamefont {Yao}},\ }\href {\doibase
  10.1007/JHEP05(2021)230} {\bibfield  {journal} {\bibinfo  {journal} {JHEP}\
  }\textbf {\bibinfo {volume} {05}},\ \bibinfo {pages} {230} (\bibinfo {year}
  {2021})},\ \Eprint {http://arxiv.org/abs/2010.01140} {arXiv:2010.01140
  [hep-ph]} \BibitemShut {NoStop}%
\end{thebibliography}%


%

\clearpage

\appendix

\section{Supplemental Material}

\section{Generalizability}
\begin{figure}[th!]
\centering
\includegraphics[width=0.48\textwidth]{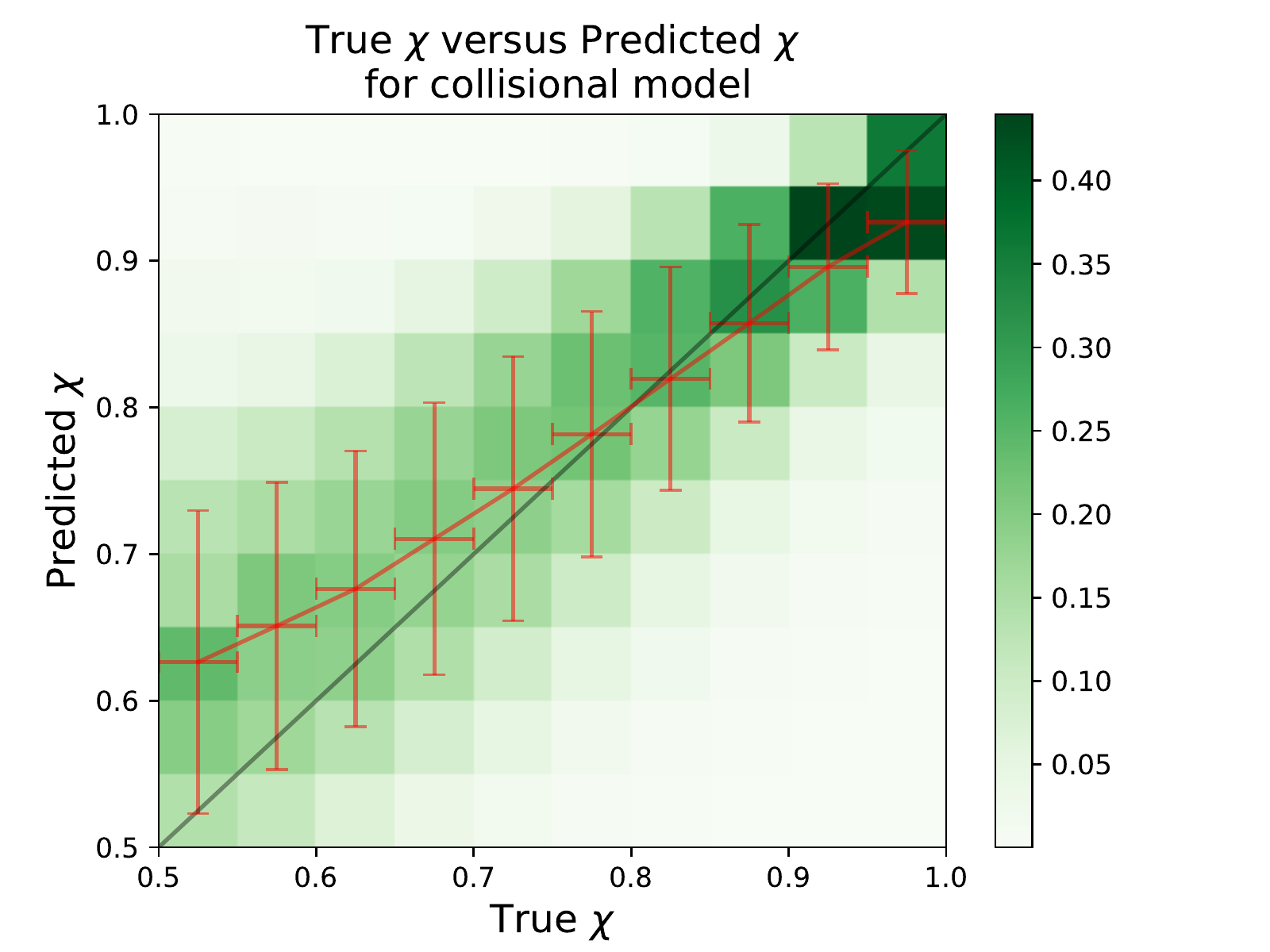}
\includegraphics[width=0.48\textwidth]{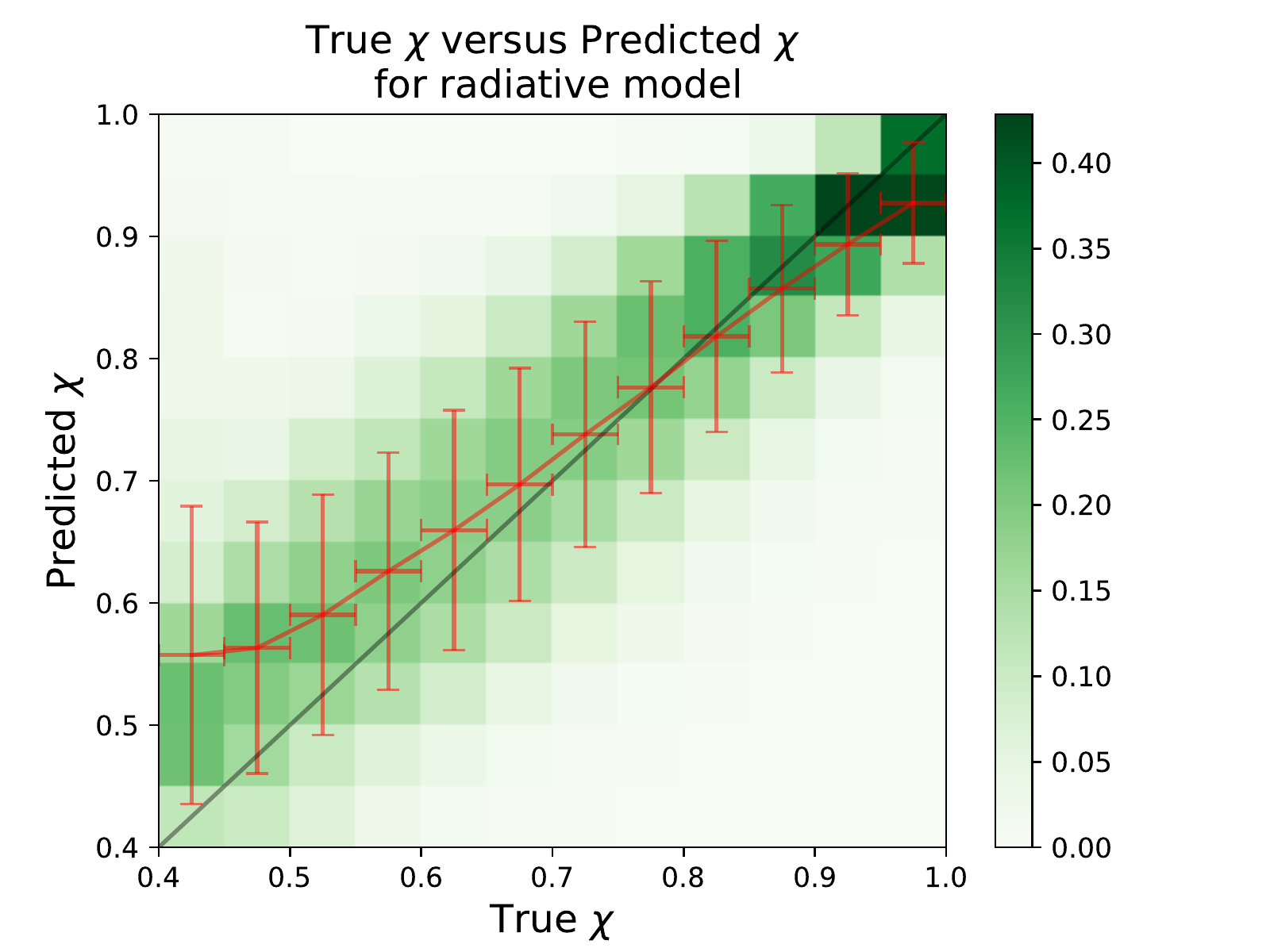}
\caption{Prediction performance of $\chi$ for collisional (upper) and radiative (lower) inspired energy loss rates within the hybrid model~\cite{casalderrey2015erratum}. The $\chi$ range differs due to the different support of the different energy loss rates, which have fairly different $\chi$ distributions. The green color represents the probability to obtain a certain value of predicted $\chi$ given a true value of $\chi$. Each column is self-normalized. The red line with error bar quantifies the average and standard deviation of the predicted $\chi$ within the given true $\chi$ bin.} 
\label{fig: Prediction_collisional}
\end{figure}

Within the hybrid strong/weak coupling model framework, we have tried prescriptions for the energy loss rate models other than the AdS/CFT one, such as the elastic/collisional and radiative inspired ones, which feature different path-length, temperature and parton energy dependence ~\cite{casalderrey2015erratum}. They serve as a preliminary verification on the generalizability of the network prediction. We show the prediction performance on both alternative energy loss prescriptions in Fig.~\ref{fig: Prediction_collisional}, which can be compared with Fig.~8 of Ref.~\cite{Du:2020pmp}. Having trained on the AdS/CFT inspired rate, one can see that the performance is still good. Further checks using a wider range of models will be done in upcoming work.

\section{Error analysis}
\begin{figure}[th!]
\centering
\includegraphics[width=0.48\textwidth]{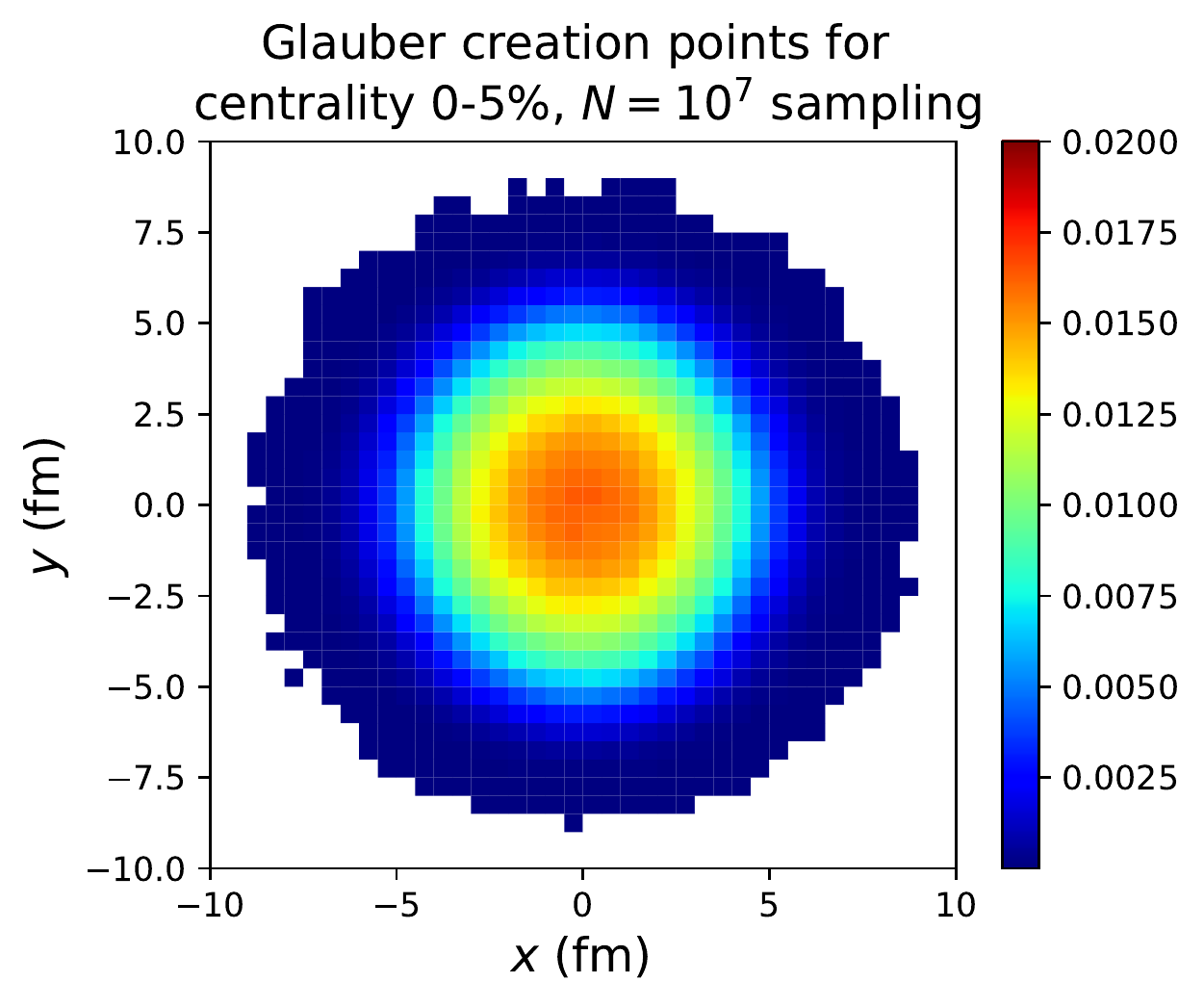}
\includegraphics[width=0.48\textwidth]{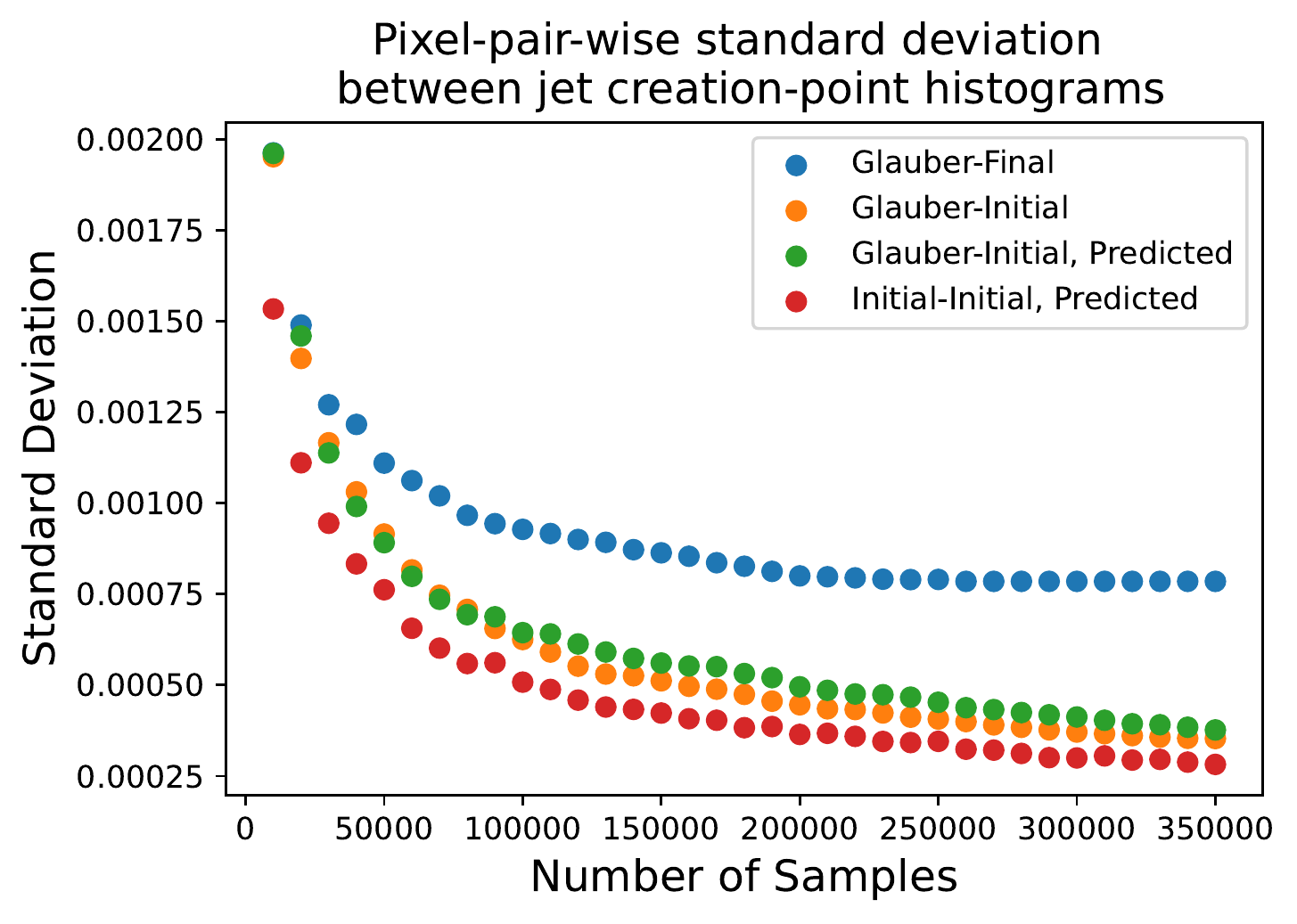}
\caption{Creation-point histogram of a genuine Glauber-like sampling for centrality 0-5\% (upper) and pixel-wise standard deviation with respect to the jet creation-point histograms for different selections as a function of the number of samples (lower).} 
\label{fig: Glauber_creation_points}
\end{figure}


The Glauber creation-point histogram with $N=10^7$ sampling are shown in the upper panel of Fig.~\ref{fig: Glauber_creation_points}. In the lower panel of Fig.~\ref{fig: Glauber_creation_points} we show the sum of the pixel-pair-wise standard deviation between the different histograms in Fig.~\ref{fig: creation points} and the Glauber one. One can see that the deviations between the Glauber, IES and IES-Predicted ones are closer, in contrast to the FES one (limited to approximately 250,000 samples), which shows that the deviation of FES is totally dominated by the surface bias physical effect and not by random noise. Deviations decrease with increasing number of jets used, limited to approximately 350,000 samples, reflecting the magnitude of the actual statistical errors (also note that our jet samples have physical weights associated with the jet $p_T$, largely reducing our effective sample number).

\begin{figure}[th!]
\centering
\includegraphics[width=0.48\textwidth]{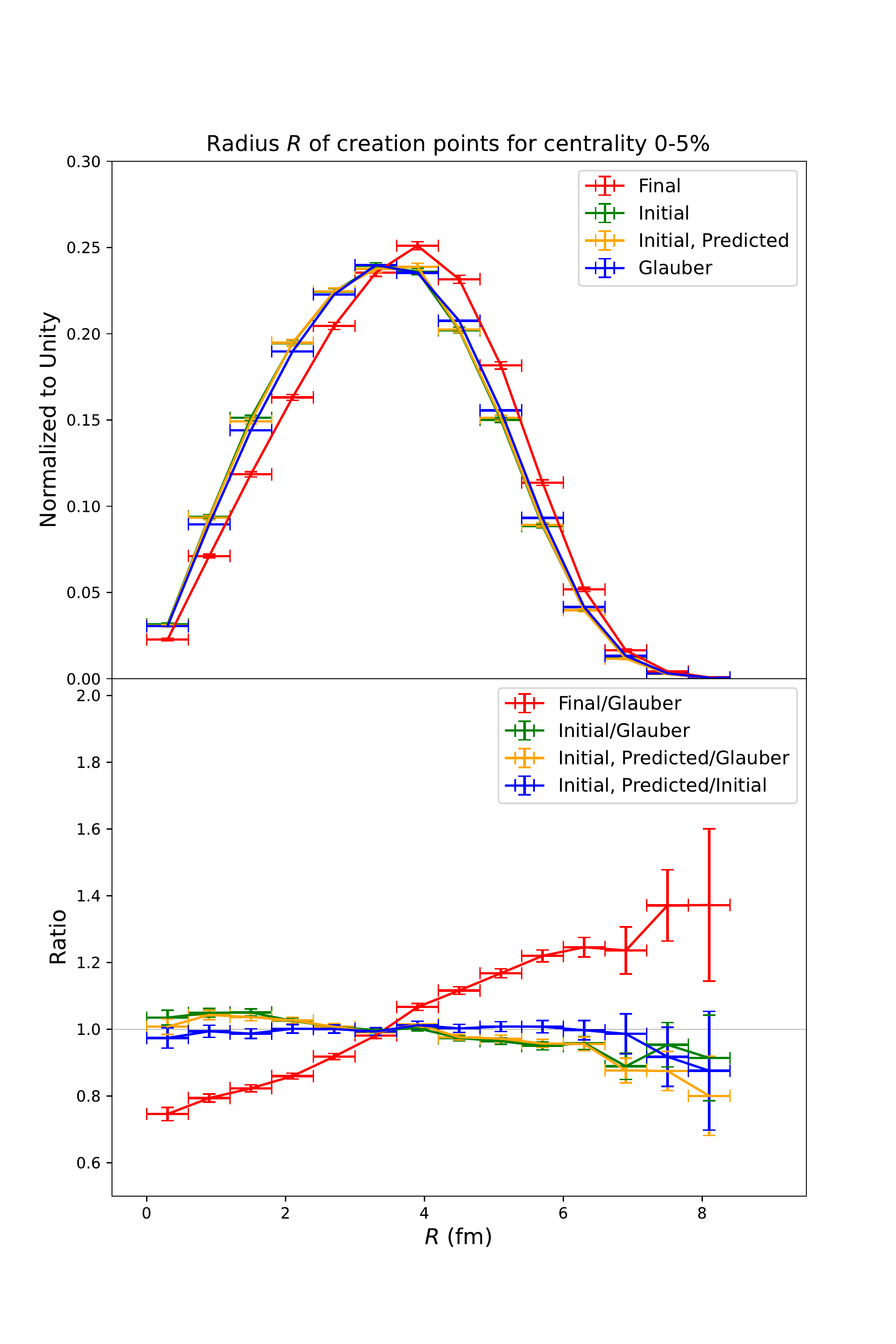}
\caption{Histograms for the radial distance of the creation point in the tranverse plane $R$ for different jet selections (upper) and some selected ratios among them (lower).} 
\label{fig: Histogram for x, y, L and Radius}
\end{figure}

Fig.~\ref{fig: Histogram for x, y, L and Radius} shows the histograms of the creation position radius $R$ for FES, IES, IES-Predicted and Glauber (upper panel), and some ratios among them (lower panel). One can see that with FES, we miss many quenched jets produced in the center of nuclear overlap region (see the red curve in the lower panel). We can see that we have slight systematic (correlated, in fact) errors in the center and periphery of the nuclear overlap region when comparing IES to Glauber (green curve of the lower panel).  Such systematic errors (associated to jet selection itself, not to the network prediction performance) most likely arise due to our initial $p_T$ cut not being sufficiently high (in a trade-off that involves jet statistics), so that a very small number of very quenched jets produced in the surface and traveling inwards fail to be counted. Indeed, the systematic errors due to the network prediction are not manifest with the current amount of jet samples (or statistics), as can be inferred from the blue curve of the lower panel.

\section{Model validation on jet $v_2$}
\begin{figure}[thb!]
\centering
\includegraphics[width=0.48\textwidth]{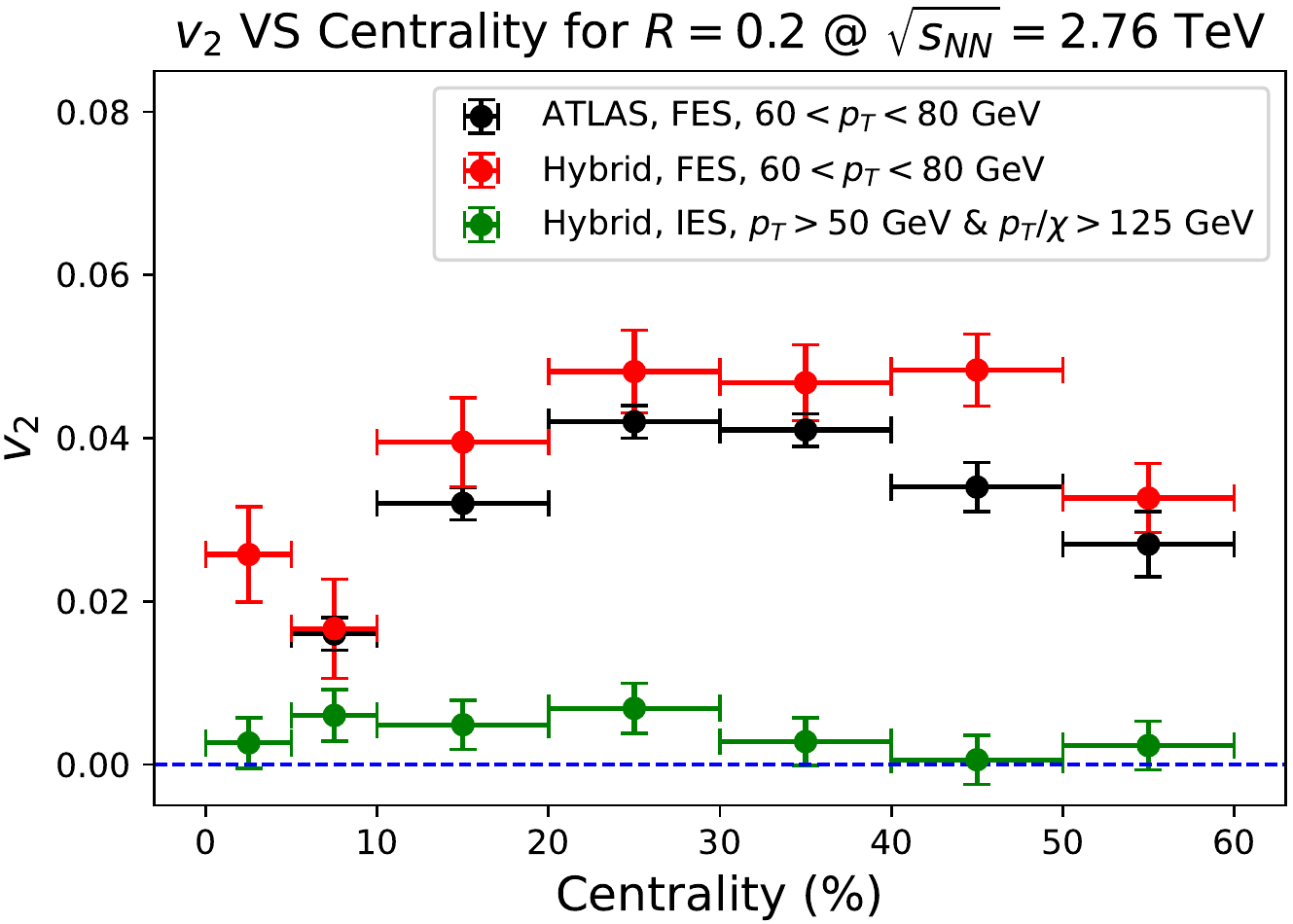}
\includegraphics[width=0.48\textwidth]{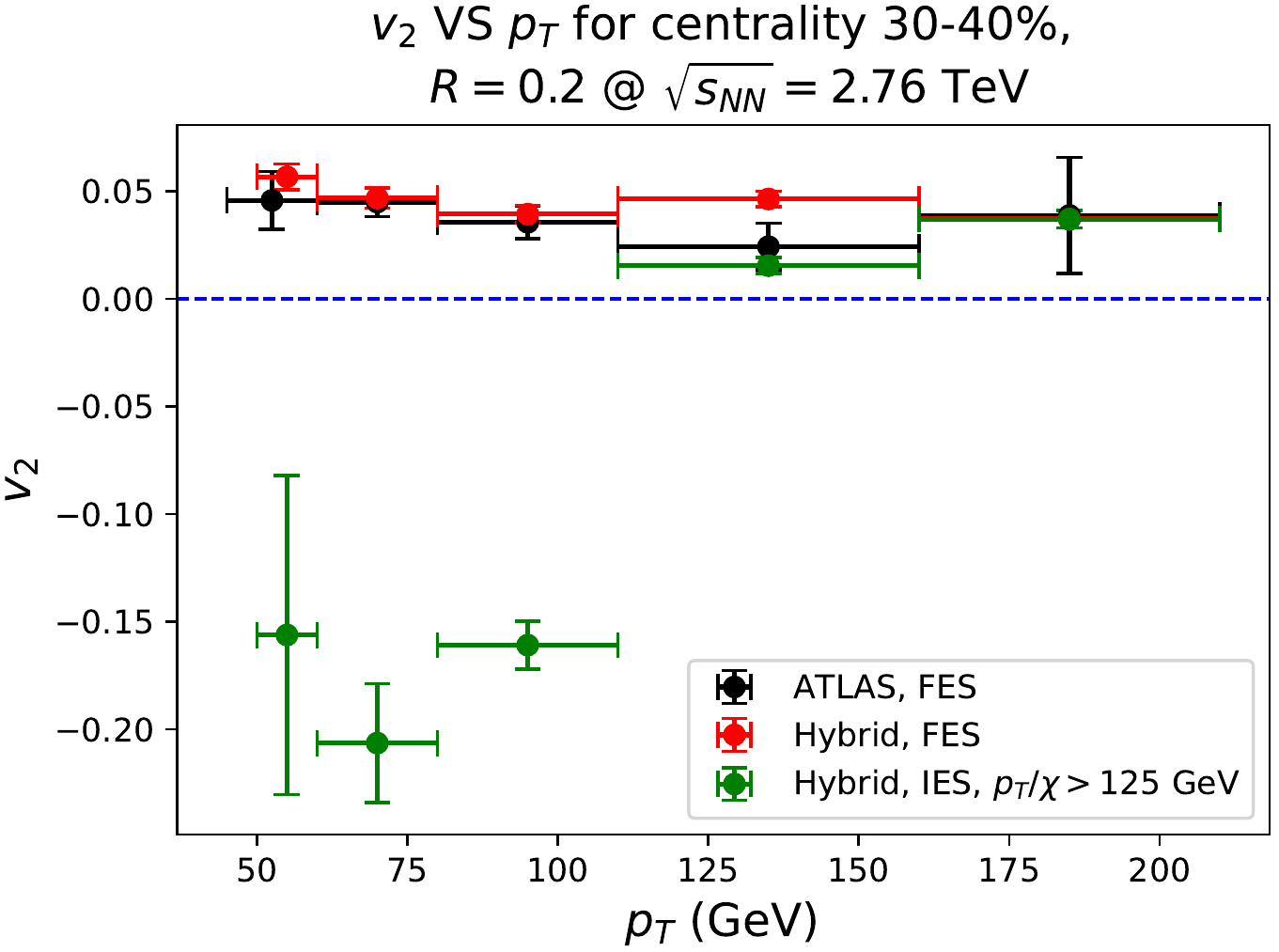}
\caption{Top: Centrality dependence of $v_2$ for FES setup (red) and IES setup with true $\chi$ (green). Bottom: $p_T$ dependence of $v_2$ for FES setup (red) and IES setup with true $\chi$ (green). Both results for the FES setup can be directly compared to ATLAS data (black)~\cite{aad2013measurement}.}
\label{fig: V2_VS_centrality_R02}
\end{figure}

We show results for jet $v_2$ in Fig.~\ref{fig: V2_VS_centrality_R02} for PbPb collisions at $\sqrt{s_{NN}}= 2.76$ TeV, using anti-$k_t$ and $R=0.2$, as a function of centrality (upper panel) and jet $p_T$ for centrality 30-40\% (lower panel). The red dots correspond to the results using FES with $60 <p_T<80$ GeV in the upper panel and $50 <p_T<210$ GeV in the lower panel, which can be directly compared to ATLAS data~\cite{aad2013measurement}, in black. Both the centrality and $p_T$ dependence of $v_2$ measured in experiments can be well described by the hybrid strong/weak coupling model. In addition, we also present the results with IES (with the cuts $\pT > 50$ GeV and $\pT/\chi > 125$ GeV) by green points. As observed in the main text, jet $v_2$ as a function of the centrality is consistent with zero (upper panel). The $\pT$-dependence of $v_2$ is quite similar to the FES when selecting jets above the employed $\pT/\chi$ cut, while the more quenched jets below this cut contribute with negative $v_2$ (lower panel).


\end{document}